\documentclass[10pt]{aastex}
\usepackage{emulateapj5,apjfonts}
\usepackage{graphicx}
\newcommand{\begit}{\begin{itemize}}
\newcommand{\enit}{\end{itemize}}
\newcommand{\begen}{\begin{enumerate}}
\newcommand{\enen}{\end{enumerate}}
\newcommand{\beq}{\begin{equation}} 
\newcommand{\eeq}{\end{equation}} 
\newcommand{\beqa}{\begin{eqnarray}} 
\newcommand{\eeqa}{\end{eqnarray}} 
\newcommand{\p}{\partial}    
 
\newcommand{\tauadv}{\tau_{\rm Adv}}
\newcommand{\tauqdotnu}{\tau_{\dot{q}_\nu}}
\newcommand{\tauqdottot}{\tau_{\dot{q}_{\rm TOT}}}
\newcommand{\brunt}{Brunt-V\"{a}is\"{a}la }

\begin{document}

\title{Viscosity \& Rotation in Core-Collapse Supernovae}

\author{Todd A. Thompson\altaffilmark{1,2}, Eliot Quataert\altaffilmark{3},
\& Adam Burrows\altaffilmark{4}}
\altaffiltext{1}{Hubble Fellow}
\altaffiltext{2}{
Astronomy Department 
\& Theoretical Astrophysics Center, 601 Campbell Hall, 
The University of California, Berkeley, CA 94720; 
thomp@astro.berkeley.edu}
\altaffiltext{3}{
Astronomy Department 
\& Theoretical Astrophysics Center, 601 Campbell Hall, 
The University of California, Berkeley, CA 94720; 
eliot@astro.berkeley.edu}
\altaffiltext{4}{Steward Observatory, 
The University of Arizona, Tucson, AZ 85721; 
burrows@as.arizona.edu}

\begin{abstract}

We construct models of core-collapse supernovae in one spatial
dimension, including rotation, angular momentum transport, and 
viscous dissipation employing an $\alpha$-prescription.  We compare the evolution of a fiducial 11\,M$_\odot$ 
non-rotating progenitor with its evolution including 
a wide range of imposed initial rotation profiles ($1.25<P_0<8$ s, where
$P_0$ is the initial, approximately solid-body, rotation period of the iron core). 
This range of $P_0$ covers the region of parameter space from where rotation
begins to modify the dynamics ($P_0\sim8$ s) to where angular velocities
at collapse approach Keplerian ($P_0\sim1$ s).
Assuming strict angular momentum conservation, all models in this range
leave behind neutron stars with spin periods $\lesssim10$ ms,
shorter than those of most radio pulsars, but similar to those expected theoretically for magnetars at birth.

A fraction of the gravitational binding energy of collapse is stored in the free energy
of differential rotation.  This energy source may be tapped by viscous processes,
providing a mechanism for energy deposition that
is not strongly coupled to the mass accretion rate through the stalled supernova shock.
This effect yields qualitatively new dynamics in models of supernovae.
We explore several potential mechanisms for viscosity in the core-collapse
environment: neutrino viscosity, turbulent viscosity caused by the magnetorotational instability (MRI), 
and turbulent viscosity by entropy- and composition-gradient-driven 
convection.   We argue that the MRI is the most effective.
We find that for rotation periods in the range $P_0\lesssim5$ s,
and a range of viscous stresses, that the post-bounce dynamics is significantly
affected by the inclusion of this extra energy deposition mechanism;
in several cases we obtain strong supernova explosions.

\end{abstract}



\section{Introduction}
\label{section:introduction}

Employing the now-standard suite of microphysics and a full solution
to the Boltzmann equation for all neutrino species, the neutrino mechanism 
of core-collapse supernovae fails  in one spatial dimension
(Rampp \& Janka 2000, 2002; Liebend\"{o}rfer et al. 2001a,b; 
Mezzacappa et al.~2001; Thompson et al.~2003; Liebend\"{o}rfer et al.~2004).
The initial "bounce shock," formed as the core density reaches $\sim$$2\times10^{14}$\,g\,cm$^{-3}$
and nuclei undergo a phase transition to free nucleons,
stalls almost immediately due to a combination of neutrino losses, the ram pressure of the infalling mantle,
and the dissociation of nuclei into neutrons and protons behind the shock.
The revival of the shock to an asymptotic energy 
of order $10^{51}$\,erg is the focus of modern supernova theory.  Any mechanism
for this revival must fundamentally rely on the transfer of gravitational binding energy to the 
post-shock mantle.   As its name makes clear, the "neutrino mechanism" (Bethe \& Wilson 1985) relies 
on neutrino interactions (primarily $\nu_e n\leftrightarrow pe^-$ and $\bar{\nu}_e p\leftrightarrow n e^+$) 
to transfer binding energy to the shocked matter in the post-bounce epoch.  
In spherical models, the stalled bounce shock remains trapped forever and continues to accrete 
the overlying  progenitor.   
Although some models show that multi-dimensional effects
might be necessary for success of the neutrino mechanism 
(Herant et al.~1994; Burrows et al.~1995; Janka \& M\"{u}ller 1995, 1996; Fryer et al.~1999; 
Fryer \& Warren 2002), multi-dimensional models employing more sophisticated neutrino 
transport fail to explode (Janka et al.~2002; Buras et al.~2003) -- albeit marginally.  

In constructing one-dimensional models, the focus of the supernova community has been 
largely on refining details of the neutrino mechanism.  These details include a fuller description
of the nuclear physics and the neutrino interactions and, over the last ten years, a precise 
handling of neutrino transport  (Mezzacappa \& Bruenn 1993a,b,c; Burrows et al.~2000; Rampp \& Janka 2000, 2002; 
Liebend\"{o}rfer et al. 2001a,b; Mezzacappa et al.~2001; Thompson et al.~2003).
Precision neutrino transport is important because the degree of neutrino heating behind the shock controls the
subsequent dynamics and because the optical depth of this region is $\lesssim0.1$. For this 
reason, the most recent one-dimensional models are fully spectral and multi-angle 
(Rampp \& Janka 2000, 2002; Liebend\"{o}rfer et al. 2001a,b; Mezzacappa et al.~2001; Thompson et al.~2003);
neutrino transport also presents the most significant technical hurdle in calculating realistic models of core collapse 
in two or three spatial dimensions (Livne et al.~2003).

It is possible that yet more refined treatments of the radiation hydrodynamics or subtle changes in the
microphysics of neutrino interactions
may yield explosions in one-dimensional models via the neutrino mechanism. 
However, it is also possible that the solution to the supernova problem does not lie in 
such technical improvements or in an accumulation of five and ten 
percent effects, but instead in new physics and order unity phenomena
not previously identified or fleshed out.  Such  improvements might include new massive pre-collapse 
progenitors with very different core entropy and lepton fraction, new 
finite temperature high-density nuclear equations of state (including 
exotic meson species, quark matter, phase transitions, etc. and the associated
neutrino microphysics), or the interplay between rotation and magnetic fields.   
In this paper, we take a step towards addressing the last of these issues in detail.  

When the core of a rotating massive star collapses, some of the gravitational binding energy of
collapse is stored in a strong shear profile.  This free energy reservoir 
may be tapped by viscous processes. In this way, the binding energy of collapse, temporarily 
trapped in differential rotation, may be transferred to the matter as thermal energy.   
This is qualitatively different from the standard neutrino mechanism, which relies
solely on neutrino interactions to communicate the binding energy of the collapsed iron core to
the rest of the star.  We find that dissipation of shear energy 
 can significantly affect the dynamics of supernovae; models that would
otherwise fail succeed as a result of viscous dissipation.  
The most promising source of viscosity is magnetic stresses generated by the magnetorotational instability (MRI; Akiyama et al.~2003)
and perhaps magneto-convection (Duncan \& Thompson 1992; Thompson \& Duncan 1993).
Although magnetic
fields in excess of $10^{15}$ G are generated, no large iron core seed field
is required for the success of this mechanism, since any seed field grows exponentially on a rotation period.

In \S\ref{section:motivate}, we provide some context and motivation for the present study and
 we review some previous research on core collapse with rotation.
In \S\ref{section:model}, we describe the technical specifications of our
one-dimensional models including rotation, viscous dissipation of shear energy, and the corresponding
transport of angular momentum.  
Section \ref{section:viscous} describes mechanisms for viscosity, including neutrino interactions, the MRI, 
and hydrodynamic convection. In \S\ref{section:rotation},
we present results from rotating models of core collapse including the centrifugal force, but without 
viscous dissipation.  We also provide a general discussion of the
conditions for explosion and an explanation for why these models and non-rotating models fail.  
We compare the magnitude of the viscous processes
discussed in \S\ref{section:viscous}  in \S\ref{section:compare}.
In \S\ref{section:dissipation}, we present results from calculations of 
core collapse including viscous dissipation.  We  show that explosions are obtained,
we explain why these models succeed, and we delineate the successful parameter space.  
In \S\ref{section:summary}, we summarize our results and discuss the implications of our findings.

\section{Scenario \& Motivation}
\label{section:motivate}

When the iron core of a massive star collapses
as a result of  the  Chandrasekhar instability, the 
implosion is reversed at nuclear densities. Here, nuclei dissociate into free nucleons 
and the equation of state stiffens dramatically, driving the bounce shock wave into the supersonic infalling outer core.
The bounce shock stalls almost immediately and a characteristic post-bounce quasi-steady-state accretion structure
is obtained.   The hot (temperatures
of order  10 MeV) newly-born protoneutron star (PNS) has a neutrinosphere radius, $R_\nu$, of 
approximately $50-80$\,km (defined here by the radius at which the optical depth to electron neutrinos $\sim$2/3,
a very energy-dependent quantity).  Overlying the PNS is a subsonic accretion flow,
bounded by the stalled stand-off bounce shock at a radius of $\sim$$150-200$ km.
The region between the PNS and the shock consists of a cooling
layer just above the PNS and a region of net heating (the "gain" region) between the cooling layer and the shock.
The boundary between net cooling and net heating is called the gain
radius, $R_{\rm g}$, typically at $\sim$100 km.  Neutrino heating is provided by both
accretion luminosity as the matter falling through the bounce shock is incorporated into
the  PNS and  core neutrino luminosity as the PNS cools and deleptonizes.

During collapse,  the inner parts of the iron core collapse to smaller radii than the outer parts.  
Thus, if one begins with a solid-body rotation profile at the moment
the core becomes unstable, as the collapse proceeds, strong differential 
rotation -- 
a negative shear ($d\Omega/d\ln r$, where $\Omega$ is the
angular velocity and $r$ is the radial coordinate) profile -- naturally develops  
(LeBlanc \& Wilson 1970; Fryer \& Heger 2000; Akiyama et al.~2003; Ott et al.~2003). 
The strong shear profile created during collapse persists as the shock is formed, as it stalls, 
and as the accretion phase begins.  The shear continues to grow, 
particularly at the PNS surface, as material is 
advected inward and the whole progenitor core is accreted.
This shear profile is a reservoir of free energy, a portion of the gravitational binding energy
of collapse trapped as differential rotation.

Heger et al.~(2000) have computed the evolution of rotating stars with zero-age-main sequence 
masses greater than $10$ M$_\odot$ and with initial equatorial surface angular velocities up to 
approximately 70\% of breakup.
Although spherical, these models account for angular momentum transport via a variety of instabilities
and include the effects of the centrifugal force up to central neon burning.  Heger et al.~(2000)
neglect rotational effects in the momentum equation and energy equation during the evolution from
neon burning to core collapse.  
The specific angular momentum they  predict
in the roughly solid-body iron core that forms prior to core collapse is $j\sim10^{16}-10^{17}$ cm$^2$ s$^{-1}$, 
with central angular velocities $\sim1-10$ rad s$^{-1}$.  On collapse, the PNS 
is formed with a mass $\sim$1.2 M$_\odot$ and a radius of $\sim8\times10^6$ cm.  
The ratio of pre-collapse to post-collapse radius for a mass element which comprises the
PNS is typically $10-20$, implying, via angular momentum conservation, an increase in 
the ratio of pre-collapse to post-collapse angular velocity of $100-400$.  Taking intermediate
values for the initial angular velocity of the iron core, the total energy
in shear can then be estimated:
\beq
E_{\rm Shear}\sim10^{52}\,\,{\rm erg}\,\,
\left(\frac{M_{\rm PNS}}{1\,\,{\rm M_\odot}}\right)
\left(\frac{R_\nu}{50\,\,{\rm km}}\right)^2
\left(\frac{\Omega}{10^3\,\,{\rm rad\,\,s^{-1}}}\right)^{2},
\label{eshear}
\eeq
corresponding to a spin period near the PNS surface of $\sim6$ ms.
The total energy in shear in the gain region, exterior to the PNS, may be less by an order
of magnitude, but is replenished on a freefall timescale as more of the progenitor falls through the standoff
supernova shock.
Although shear energy in the range of $10^{51}-10^{52}$ erg is small on the scale of the binding energy
of the resulting fully contracted neutron star and the energy liberated in neutrinos, it is large on
the scale of the canonical asymptotic supernova energy and (not coincidentally)
the binding energy of the matter exterior to the gain radius, $\sim10^{51}$ erg.  If this free energy
source can be tapped on a timescale comparable to, for example, the neutrino heating
timescale in the gain region, we expect large modifications to the dynamics.

Any local viscous process acting in a region of differential rotation will transport angular momentum
and dissipate the energy stored in shear on a viscous timescale.  It is this
mechanism for transferring gravitational binding energy, stored during collapse as shear energy,
to the matter in the gain region that we explore in this paper.  
To this end, we consider several potential mechanisms for viscosity: microscopic shear 
neutrino viscosity and the turbulent viscosity caused by convective or magnetic stresses.
Generation of these turbulent stresses is caused by unstable entropy, composition, or angular velocity gradients.
As an expedient, we distinguish between magnetic stresses caused by the 
magnetorotational instability (MRI) and the turbulent viscosity of purely hydrodynamical convection (see \S\ref{section:mri}
and \S\ref{section:convection}).
We compare these agents of viscosity and find that the MRI and hydrodynamical convection
dominate microscopic neutrino viscosity and that the latter is likely not capable of modifying the
dynamics of supernovae.  
We find that for fast enough initial rotation periods, dissipation of shear energy
can affect the subsequent dynamics significantly; a stalled shock can be revived and driven to infinity;
a dud is transformed into an explosion.   We emphasize that for the success of this mechanism we
do not require large seed magnetic fields in the iron core prior to collapse.  The large magnetic
fields generated from even a vanishingly small initial field are a necessary consequence 
of the MRI in rapidly rotating collapse (Akiyama et al.~2003).

The calculations presented in this paper are one-dimensional. Spherical symmetry is 
assumed here, but demonstrably broken in a
rotating system.   We have attempted to partially mitigate this issue by restricting our calculations to those in which,
throughout the evolution, Keplerian rotational velocities are not attained.  That is, for every model presented,
at every time and at all radii, $\Omega(r)<\Omega_{\rm K}=\sqrt{GM/r^3}$.  By imposing this restriction,
the momentum equation is dominated by the gradients in the thermal pressure and the gravitational force; 
we calculate stars, not disks.  That said, for $E_{\rm shear}$ to be in an interesting energy range (say $\gtrsim10^{50}$ erg)
rotation rates for the iron core at collapse must be in the range of a few rad s$^{-1}$.  Assuming strict
conservation of angular momentum and that the supernova leaves behind a $\sim1.4$ M$_\odot$ neutron
star with a radius of $\sim10$ km, we find that this energy requirement implies an initial spin period for the
young neutron star of at most $\sim$10 milliseconds.  Observations of radio pulsars imply that the initial
rotation rate of neutron stars should be in the range of many 
tens to hundreds of milliseconds (e.g.~Kaspi \& Helfand 2002), factors
of hundreds to tens of thousands lower in rotational energy.  Thus, although we restrict ourselves
to models that are slowly rotating on the scale of $\Omega_{\rm K}$, they are rapidly rotating
on the scale of observed neutron stars.  However, caution is warranted in taking this argument too seriously.
Gravitational radiation (e.g.~Ostriker \& Gunn 1969),  magnetic dipole radiation 
(e.g.~Pacini 1967, 1968; Gunn \& Ostriker 1969; Lindblom et al.~2001; Arras et al.~2003), 
magneto-centrifugal winds (Thompson 2003; 
Thompson, Chang, \& Quataert 2004), and late-time fallback 
(see Woosley \& Heger  2003 and references therein) can spindown young neutron stars,
complicating the inference of "initial" neutron star spin periods (Kaspi \& Helfand 2002).
Nevertheless, it is possible that rapid rotation is relevant in only a subset of all core-collapse supernovae
(e.g.~those producing magnetars; Duncan \& Thompson 1992; Thompson \& Duncan 1993).
The size of that subset is unknown.

In addition to the various processes that may slow a neutron star at or just after birth, it is also possible
that stellar evolution produces slowly rotating iron cores.
Recent models of Heger et al.~(2003) and Heger et al.~(2004), based on the work of Spruit (2002, 2003), attempt to evolve core-collapse
progenitors including angular momentum transport via magnetic processes.  
These effects yield significantly smaller iron core specific angular momenta
($j\sim10^{14}-10^{15}$ cm$^2$ s$^{-1}$, $\Omega\lesssim0.1$ rad s$^{-1}$), implying much lower shear energy.
Even smaller rotation periods were found in Spruit \& Phinney (1998).
Hence, uncertainty in models of rotational stellar evolution remains high.  

Similar uncertainty surrounds models of rotating core collapse and the role of rotation and
magnetic fields in the mechanism of supernovae.
Although  pioneering, the MHD calculation of rotational core-collapse  by
Leblanc \& Wilson (1970) is now largely obsolete because the neutrino microphysics, nuclear equation
of state, and progenitor configuration they employed have been superseded.  
The jet-like explosion their calculation produced was reconsidered  with more modern input
physics  by Symbalisty (1984), who showed that only with 
initial iron core spin periods much less than one second and a very large fossil seed field in the
progenitor iron core, could jet-driven explosions be obtained.

The recent work of Akiyama et al.~(2003) considered the potential effects of the MRI in stellar collapse.
They  showed that much of the post-bounce core collapse environment 
is unstable to the MRI and that large saturation magnetic fields are naturally produced. 
Energy deposition via local turbulent dissipation was neglected in their study.
Instead, Akiyama et al.~(2003) argued that the very large field strengths generated by the MRI create a jet 
with sufficient MHD luminosity to explode the star as in the models of Khokhlov et al.~(1999).  This argument assumes that
the magnetic field generated by the MRI can form the organized large-scale
fields required for collimation and jet formation. 
A limitation of the models presented by Akiyama et al.~(2003) is that they were not self-consistent.
They did not include rotation in the dynamics and because they neglected angular momentum transport, 
the energy available for the MHD jets they consider may have been overestimated.

Fryer \& Heger (2000) considered rotating core-collapse in two spatial dimensions, and recently
Fryer \& Warren (2004) extended these rotating models to three dimensions.
The only effect of rotation on the explosion mechanism 
that Fryer \& Heger (2000)  and Fryer \& Warren (2004) discuss is the suppression of 
convection in the equatorial plane due to the stabilizing centrifugal force.
However, for stabilization to
occur, rotation must be important enough that if the calculation had been performed
in MHD the equatorial region would have been destabilized by the MRI.   Thus, the stabilization 
that occurred  in their hydrodynamic calculations 
is somewhat artificial and it is a signal that the simulations must be performed in MHD.
All models computed in Fryer \& Heger (2000)  and Fryer \& Warren (2004) explode.  For this reason, 
Fryer \& Heger (2000)  and Fryer \& Warren (2004) did not assess
the role of viscous heating, the competition between different possible
sources of viscosity, or the potential role of rotation itself in generating explosions.
Finally, Fryer \& Warren (2004) concluded
that the saturation magnetic field strength throughout the post-collapse structure is
orders of magnitude lower than the field estimated in Akiyama et al.~(2003).
We address this conclusion in  \S\ref{section:rotation} and find that large fields 
in the range $10^{14}-10^{16}$ G can be  typical in rotating core collapse (
Duncan \& Thompson 1992; Thompson \& Duncan 1993; in agreement with Akiyama et al.~2003).

\section{The Model}
\label{section:model}

The core-collapse supernova code we have developed and that we employ
here, SESAME\footnote{{\bf S}pherical {\bf E}xplicit/Implicit {\bf S}upernova 
{\bf A}lgorithm for {\bf M}ulti-Group/Multi-Angle {\bf E}xplosion Simulations.},  is 
described in detail in Burrows et al.~(2000) and Thompson et al.~(2003).
The hydrodynamics scheme is Lagrangean, Newtonian, explicit,
and employs artificial viscosity for shock resolution.
The radiation transport algorithm solves the order $V/c$
Boltzmann equation (Eastman \& Pinto 1993) for three neutrino species: $\nu_e$, $\bar{\nu}_e$,
and "$\nu_\mu$", employing the standard approximation 
that the non-electron neutrino species can be treated collectively.
It is fully spectral, employs the tangent-ray technique for angular discretization of
the radiation field, and is based on the Feautrier variables.
The neutrino physics employed is standard and described
in Bruenn (1985), Thompson (2002), Burrows \& Thompson (2002), and Thompson et al.~(2003).
Notable additions include nucleon-nucleon bremsstrahlung (Burrows et al.~2000; Thompson et al.~2000),
and weak-magnetism and recoil corrections to both the charged-current neutrino-nucleon absorption/emission
processes and elastic neutrino-nucleon scattering (Horowitz 1997; Horowitz 2002).
Inelastic neutrino-electron/positron scattering is calculated with a fast and accurate
explicit scheme described in Thompson et al.~(2003).
The equations of radiation hydrodynamics are coupled to an efficient
tabular version of the Lattimer-Swesty
high-density nuclear equation of state (for $\rho\gtrsim10^7$\,g\,cm$^{-3}$) 
and the Helmholtz equation of state (for $\rho\lesssim10^7$\,g\,cm$^{-3}$;
Timmes \& Arnett 1999; Timmes \& Swesty 2000; Thompson et al.~2003).\footnote{
A recent comparison between the Lattimer-Swesty (LS)EOS in NSE at moderate 
densities and temperatures (characteristic of the
gain region in models of supernovae) with other NSE EOSs indicates a discrepancy of potential importance.  
Free nucleon and $\alpha$-particle number fractions as well as entropy profiles disagree at the $\sim10-30$\%
level in some regions of thermodynamic space. The comparison is documented by F.~X.~Timmes 
online at {\it http://www.cococubed.com/code\_pages/eos\_supernova.shtml}. 
Janka et al.~(2004) have compared supernova models computed
with the LSEOS with otherwise identical models computed with the
high-density EOS of Shen et al.~(1998).  They find quantitative 
differences between the two models, but they do not find that
those differences lead to qualitatively different model outcomes;
both models fail to yield explosions.  Because these models do not
show large qualitative differences and because in this paper  
we make a relative comparison between models without rotation, 
with rotation, and with viscosity and angular momentum transport,
we have employed the standard LSEOS.}
All of the models presented here are calculated with 20 energy groups per
neutrino species and 500 Lagrangean mass zones.

\subsection{Additions to the Standard Algorithm}

To the standard set of Newtonian equations for Lagrangean hydrodynamics
in one spatial dimension, we add rotation via the term 
$f_\theta j^2/r^3$ in  the momentum equation, where $j$ is the specific angular momentum,
$r$ is the radial coordinate, and $f_\theta(=2/3)$ results from 
taking the angular average  for one dimensional parameterizations of rotation (Heger  et al.~2000).

Angular momentum transport is calculated by solving the equation
\beq
\frac{Dj}{Dt}=\frac{1}{\rho r^2}\frac{d}{dr}\left(
\xi\rho r^4\frac{d\Omega}{dr}\right),
\label{ang}
\eeq
where $D/Dt=\p/\p t + {\bf V}$\mbox{\boldmath $\cdot\nabla$}, $\rho$ is the mass density, 
$\Omega$ is the angular velocity, and $\xi$ is the shear viscosity.\footnote{Our use of ``$\xi$" 
to denote ``viscosity" alleviates confusion with ``$\nu$", which we use to denote neutrinos or neutrino
quantities.} 
Equation (\ref{ang}) is solved explicitly in operator-split fashion 
from the rest of the radiation-hydrodynamical algorithm.  
The calculations presented in this paper conserve total angular momentum at the $\sim0.05\%$ level;
maximum fractional deviations in the total angular momentum do not exceed $\sim0.15\%$.
During a given calculation, depending 
on the magnitude of the viscosity, the diffusive timestep 
($\Delta t<{\rm min}[(\Delta r)^2/2\xi]$) becomes shorter than the Courant-limited 
hydrodynamical timestep and we 
sub-cycle the calculation of the update to $j$.

The dissipation of shear energy is handled by adding a term to
our Lagrangean equation for the specific internal energy ($e$);
\beq
\frac{De}{Dt}=\,[\dots]\,+\,\xi\left(\frac{d\Omega}{d\ln r}\right)^2,
\label{dedt}
\eeq
where the actual form of $\xi$ and the resulting viscous energy deposition rate
($\dot{q}_{\xi}$) depends on the specific viscous process considered.
Note that we dissipate shear energy locally.  For a source of microscopic viscosity 
this is appropriate.  However, for a turbulent viscosity like the MRI,
it may be that energy stored in the turbulent magnetic field can be
transported and deposited non-locally.  This would relieve the requirement
in the models presented here that a region of high local viscous dissipation
must correspond to a region of large shear.  However, a multi-dimensional
MHD treatment is required to address this question fully in the supernova context.

\subsection{The Progenitors}
\label{section:progenitor}

We consider two progenitors.  The first is the non-rotating 11\,M$_\odot$ progenitor
of Woosley \& Weaver (1995).  This model is zoned out to 1.45 M$_\odot$, corresponding
to a radius of $\sim4000$ km.  We impose  rotation on the core of the progenitor, taking
\beq
\Omega(r)=\frac{\Omega_0}{1+(r/R_\Omega)^2},
\label{omegaprofile}
\eeq
where $r$ is the spherical radial coordinate and not the cylindrical radial coordinate.
Equation (\ref{omegaprofile}) implies that the rotation profile is roughly solid-body out to $R_\Omega$.
Comparing with the rotating stellar progenitors of Heger et al.~(2000),
eq.~(\ref{omegaprofile}) provides a reasonable approximation to $\Omega(r)$
with $R_\Omega$ near $\sim1000$ km.
Our models  span spin periods  in the range $1\lesssim P_0\lesssim 10$ seconds
($P_0=2\pi/\Omega_0$). 
As discussed in \S\ref{section:motivate}, we restrict ourselves to models in which
$\Omega(r)<\Omega_{\rm K}$ at all times and at all radii during a given calculation.
For our 11 M$_\odot$ model and for $\Omega(r)$ set by eq.~(\ref{omegaprofile}) with $R_\Omega=1000$ km,
we find that $P_0\ge1.25$ s satisfies this constraint.  This sets our lower bound on $P_0$.
The upper bound on $P_0$ is set by the fact that our spherical models with $P_0>8$ s and $R_\Omega=1000$ km are nearly
indistinguishable from models with $P_0=\infty$.  In all models with this 11 M$_\odot$
progenitor we take $R_{\Omega}=1000$ km.

The second progenitor we consider is the rotating 15 M$_\odot$ model E15A of Heger et al.~(2000),
which we zone out to an enclosed mass of 1.65 M$_\odot$.  With this model we can compare
directly to the 2D axisymmetric core-collapse calculations of Fryer \& Heger (2000)
and the 3D calculations of  Fryer \& Warren (2004).   

Figure \ref{plot:omegap2} shows $\Omega(r)$ for the progenitor models considered here.
Solid lines are for the 11 M$_\odot$ progenitor with $P_0=1.25$, 2, 3, 4, 5,  and 8 seconds.
The dashed line shows $\Omega(r)$ for the 15 M$_\odot$ model from Heger et al.~(2000).
Throughout this paper we use "E15A" to refer to the 15 M$_\odot$ progenitor of Heger et al.~(2000)
and  "$P_0=$" in referring to the 11 M$_\odot$ models.

\section{Viscosity}
\label{section:viscous}

In this section, we discuss the pure microscopic shear viscosity caused by neutrinos
and two forms of turbulent viscosity: the magnetorotational instability and
hydrodynamic convection.  We compare the magnitude of these
viscous processes in \S\ref{section:compare} in the context of the rotating models
of core collapse presented in \S\ref{section:rotation}.  In \S\ref{section:dissipation},
our results for rotating models including viscous dissipation are discussed.
 
\subsection{Neutrino Viscosity}
\label{section:neutrino}

When neutrinos are fully diffusive, the shear viscosity is given by (van den Horn \& van Weert 1984;
Burrows \& Lattimer 1988; Thompson \& Duncan 1993)
\beq
\xi_\nu=\frac{4}{15}\frac{E_\nu}{\rho c}\langle\lambda_\nu\rangle,
\label{nunu}
\eeq
where $E_\nu$ is the neutrino energy density, 
$\langle\lambda_\nu\rangle$ is the average neutrino mean-free path,
\beq
\langle\lambda_\nu\rangle=\left\{\int d\varepsilon_\nu 
\varepsilon_\nu^2 \lambda_\nu(\varepsilon_\nu)\,\tilde{J}_\nu(\varepsilon_\nu)\right\}/
\left\{\int d\varepsilon_\nu 
\varepsilon_\nu^2 \,\tilde{J}_\nu(\varepsilon_\nu)\right\},
\eeq
\beq
\tilde{J}_\nu(\varepsilon_\nu)=\frac{1}{2}\int_{-1}^{1}d\mu\,f_\nu(\mu,\varepsilon_\nu),
\eeq
$f_\nu$ is the invariant neutrino phase-space distribution function,
$\varepsilon_\nu$ is the neutrino energy, and $\mu=\cos\theta$.

Such a prescription for the viscous effects of neutrinos applies
only in the region of diffusive neutrino transport, below the neutrinospheres,
where $\langle\lambda_\nu\rangle \ll r$.  In the optically thin region,
as $\langle\lambda_\nu\rangle\rightarrow\infty$, eq.~(\ref{nunu}) is inapplicable.  Nearest the neutrinospheres, $\xi_\nu$ will
be large, but outside this region the viscosity is not well defined.
However, the effect of the streaming neutrinos interacting with the background
differential rotation profile and the distribution of hydrodynamic (convective)
and magnetohydrodynamic (magneto-convective) fluctuations can still be quantified.
Neutrinos propagating through the cooling and gain regions damp
fluctuations with a characteristic rate
(Jedamzik et al.~1998; Agol \& Krolik 1998)
\beq
\Gamma_{\nu} \sim \frac{E_\nu}{\langle\lambda_\nu\rangle\rho c}.
\label{gammanunu}
\eeq
For example, if a convective mode of wave number $k$ in the gain region has linear growth rate given
$\sqrt{|N^2|}$, then if $\Gamma_{\nu} > \sqrt{|N^2|}$ the 
convective mode will be damped.  
Similarly, if the MRI grows in the gain region
on a timescale $\sim\Omega^{-1}$, if $\Gamma_{\nu}>\Omega$ the linear MRI mode will also be damped.

\subsection{The Magnetorotational Instability}
\label{section:mri}

In the equatorial region, neglecting gradients in the $\theta$-direction\footnote{The Solberg-H\o iland
criterion for instability contains latitudinal gradients that for a star may be as large
as  those in the  radial direction.  For the purposes of the discussion here,
and in the spirit of the one-dimensional calculations presented in this paper, we retain only
the latter.}, the Solberg-H\o iland condition for instability in the presence of rotation, but without magnetic fields is given by (Tassoul 1978)
\beq
N^2+\kappa^2<0,
\label{so}
\eeq
where $\kappa$ is the epicyclic frequency,
\beq
\kappa^2=4\Omega^2+\frac{d\Omega^2}{d\ln r},
\eeq
and $N$ is the \brunt frequency.  In the supernova context, 
material may be driven convectively unstable by negative gradients
in entropy ($s$)  or in lepton number ($Y_l$). For our purposes here, it is sufficient to replace
$Y_l$ with the electron fraction ($Y_e$),  thereby neglecting the electron 
neutrino fraction $Y_{\nu_e}$ in regions where this species is trapped and diffusive.  
The expression for the \brunt frequency
used in this paper is then (Lattimer \& Mazurek 1981)
\beq
N^2=\frac{g}{\gamma}\left(
\frac{1}{P}\left.\frac{\p P}{\p s}\right|_{\rho,Y_e}\frac{ds}{dr}+
\frac{1}{P}\left.\frac{\p P}{\p Y_e}\right|_{\rho,s}\frac{dY_e}{dr}\right),
\eeq
where $g$ is the effective gravity and $\gamma=d\ln P/d\ln\rho|_s$.  
When magnetic fields are included in the stability analysis of differentially
rotating bodies, a "generalized" Solberg-H\o iland criterion is obtained
(Balbus \& Hawley 1991; 1992a,b; 1994; 1998).  The condition
\beq
N^2+\frac{d\Omega^2}{d\ln r}<0,
\label{gso}
\eeq
replaces eq.~(\ref{so}) as the criterion for instability in the equatorial region.  Ignoring gradients
in entropy and composition, a magnetized differentially
rotating fluid is unstable to the magnetorotational instability if $d\Omega^2/d\ln r<0$.
The linear growth rate for the fastest  growing mode is (Balbus \& Hawley 1991)
\beq
\Gamma_{\rm MRI}=\frac{1}{2}\left|\frac{d\Omega}{d\ln r}\right|.
\label{gammamri}
\eeq
To rough, but useful, approximation, $\Gamma_{\rm MRI}\sim\Omega$.
In a sub-sonic, steady-state accretion flow with net heating, as in the gain region just
interior to the shock in the supernova context, the entropy gradient is negative, and
compositional gradients are insufficient to stabilize this region.  Because the process of collapse,
along with angular momentum conservation implies negative shear in the gain region,
eq.~(\ref{gso}) implies that as long as $\Gamma_{\nu}<\Gamma_{{\rm MRI}}$
this region must be unstable to the MRI.

Our form for the viscosity due to the MRI is motivated by consideration of  the Maxwell
stress ($t_{r\phi}$); 
\beq
t_{r\phi}=B_rB_\phi/4\pi\sim B^2/4\pi=V_A^2\rho,
\eeq
where $V_A=B/\sqrt{4\pi\rho}$ is the Alfv\'{e}n velocity.
The viscous shear stress is given by $\xi\rho |d\Omega/d\ln r|\sim\xi\rho\Omega$.
Equating the Maxwell stress and the shear stress we find that $\xi_{\rm MRI}\sim V_A ^2/\Omega$. 
For the MRI, $V_A k \sim \Omega$ for the fastest growing mode of wave number $k$.
Taking $k\sim1/H$, where $H$ is the pressure scale height, and substituting, we find that
$\xi_{\rm MRI}\sim H^2\Omega$, up to factors of $d\ln\Omega/d\ln r$.  We introduce a 
constant of order unity, an $\alpha$ parameter (typically 0.1), and take
\beq
\xi_{\rm MRI}=\alpha H^2\Omega.
\label{numri}
\eeq
Equation (\ref{numri}) is appropriate for the form of the viscous stress 
regardless of the ratio $V_\phi/c_s$.\footnote{Akiyama et al.~(2003) 
estimate the MRI viscous timescale as $\tau_{\rm Visc}\sim B_\phi\beta_{\rm M}/(2B_r\Omega)$,
where $\beta_{\rm M}=P/(B^2/8\pi)$ (see their eq.~32).  Because $\beta_{\rm M}\gg1$ in a star, they neglect
the redistribution of angular momentum by magnetic stresses.  This is incorrect.  In estimating the  $\alpha$ parameter, they take $\alpha\sim B_rB_\phi/(4\pi P)$,
and it is in this way that $\beta_{\rm M}$ enters their $\tau_{\rm Visc}$.  Their estimate for $\alpha$,
however, is relevant for disks only.  In the stellar context,
$\tau_{\rm Visc}\sim(\alpha\Omega)^{-1}(r/H)^2$.  
For most of our models this estimate yields $\tau_{\rm Visc}\sim0.1$ s at $R_\nu$ and $\tau_{\rm Visc}\sim1$ s 
at $R_{\rm g}$  and so
angular momentum redistribution cannot be neglected.
}
In all models considered in this work, $V_\phi$ is less than or much less than $c_s$ and $V_{\rm Kep}$.
This condition emphasizes the fact that we consider stars, not disks.  In accretion disks,
$H=c_s/\Omega$.  Substituting into eq.~(\ref{numri}) one obtains the classic form for the 
shear stress in accretion disks from Shakura \& Sunyaev (1973), $\xi=\alpha c_s H$.
Another way to see that eq.~(\ref{numri}) is the appropriate form for the viscosity is to recognize that
any turbulent viscosity can be estimated as the product of a turbulent velocity ($V_{\rm turb}$) 
and a correlation length ($L$) for fluctuations.  
In the case of a disk the adiabatic sound speed $c_s$ is $V_{\rm turb}$ and
$L$ is the pressure scale height, $H$. In a star, with $H\Omega$ less than or {\it much} less than $c_s$,
the shear across a scale height $H\Omega$ takes the place of $c_s$ as $V_{\rm turb}$.  The correlation length for the MRI 
remains the pressure scale height and eq.~(\ref{numri}) follows.

Beginning with an arbitrarily small seed field, the MRI increases the magnetic field strength
exponentially on a timescale $\sim\Omega^{-1}$.  In models of accretion disks, the MRI in 
ideal MHD saturates such that the toroidal component of the Alfv\'{e}n speed comes into
rough equipartition with the sound speed.  In the stellar context considered here,
the Alfv\'{e}n speed should saturate at of order the rotation speed,
\beq
B_{\rm Sat_\phi}\sim(4\pi\rho)^{1/2}\,V_\phi.
\label{bsat}
\eeq
The value of the toroidal saturation magnetic energy density ($B_{{\rm Sat}_\phi}^2/8\pi$)
may be somewhat less (a factor of $\sim$10) than the value implied by 
eq.~(\ref{bsat}) (see, e.g., Hawley et al.~1996 and Stone et al.~1996  for results in the accretion context).   
A reduction of this magnitude in
the energy density implies a rather small correction to the field strength $B_{{\rm Sat}_\phi}$.
Further possible corrections to eq.~(\ref{bsat}) are order unity and include factors of $d\ln \Omega/d\ln r$
(Akiyama et al.~2003).
The magnitude of the saturation energy density in the poloidal direction 
is roughly an order of magnitude smaller than $B_{\rm Sat_\phi}$ (e.g.~Balbus \& Hawley 1998).  

\subsection{Convective Viscosity}
\label{section:convection}

The \brunt frequency is always negative in the gain region. Convection,
driven by neutrino heating, results (Herant et al.~1994;
Burrows, Hayes, \& Fryxell 1995; Janka \& M\"{u}ller 1996; Mezzacappa et al.~1998; Buras et al.~2003).
If convection has time to develop fully, and the Reynolds number is high, 
the presence of turbulent convection may result in the dissipation 
of shear energy in much the same way as turbulence caused 
by the MRI.
The convective viscosity ($\xi_{\rm Con}$) can be naively estimated as
the product of the turbulent velocity (in this case the convective velocity $V_{\rm Con}$)
and the correlation length:
\beq
\xi_{\rm Con}\sim H V_{\rm Con}/3.
\label{nucon}
\eeq
The convective velocity can be estimated from mixing length theory:
\beq
V_{\rm Con}=2gl(\delta\rho/\rho),
\eeq
where $g=-\rho^{-1}dP/dr=P/\rho H$, $l$ is the mixing length,
and $\delta\rho=(d\rho/dr-\Delta\rho/l)l$.  In the gain region, $V_{\rm Con}$
is typically $\sim5\times10^{8}-10^9$ cm s$^{-1}$. 
Unfortunately, this estimate is very rough because even the sign of the Reynolds stress, which dictates the direction
of angular momentum transport in a convective region, is uncertain
and may depend sensitively  on the Rossby number, the ratio of the rotation period
to the convective turnover time (Chan 2001; K\"{a}pyl\"{a}, Korpi, \& Tuominen 2003).
In general, for Rossby numbers less than unity the Reynolds stress is positive and
angular momentum transport is outward. However, this statement depends on latitude
and the magnitude of the heat flux through the convection zone.
In the supernova context, particularly for rotation rates with $V_\phi\ll c_s$ in the
gain region (and deep in the PNS interior), we expect the Rossby number to be greater than unity 
(Thompson \& Duncan 1993).

It is not strictly correct to speak of two different
mechanisms for  turbulent convective viscosity; the MRI and hydrodynamical convection come
together.  There is just one turbulent stress tensor, containing Reynolds and Maxwell
correlations, of the form 
$\langle u_r u_\phi- u_{A_r}u_{A_\phi}\rangle$, where $u$ is the fluctuation velocity and 
$u_{A_i}=B_i/\sqrt{4\pi\rho}$ is the fluctuation Alfv\'{e}n speed in direction $i$;
and convection itself will, of course, amplify magnetic fields even in the absence of the MRI,
thereby generating appreciable magnetic stresses (Duncan \& Thompson 1992; Thompson \& Duncan 1993).
But one term in the stress tensor may dominate the other and for the 
purposes of this paper we distinguish between purely hydrodynamical convection 
($\Gamma_{\rm Con}\gg\Gamma_{\rm MRI}$) and the case where the Maxwell stress
is comparable to or dominates the Reynolds stress (see Narayan et al.~2002
for a more general discussion).

\section{Core Collapse with Rotation, but without Viscous Dissipation}
\label{section:rotation}

In this section we describe results from a set of simulations with the 11 M$_\odot$ 
progenitor with $P_0=1.25$, 2, 3, 4, 5, and 8 seconds as well as results with model E15A from Heger et al.~(2000)
(see \S\ref{section:progenitor}).  In these models we include the centrifugal force, but neglect
diffusive angular momentum transport.  We do this in order to separate the effects of the centrifugal
force from the effects of viscous dissipation.
Table \ref{tab:rot} summarizes a few properties
of these rotating models.  We list the model
names, the shear energy at bounce and 500 ms after bounce, and the ratio of total rotational
energy to total gravitational energy ($\beta^{\rm Rot}$) initially, at bounce, and 500 ms after bounce.
In \S\ref{section:dissipation}, we describe results from these models when viscous dissipation is included.  

Figure \ref{plot:ltet} shows the neutrino luminosity (left panels, in units of $10^{52}$\,erg\,s$^{-1}$) and
average rms neutrino energy (right panels, in MeV) at infinity as a function of time relative to
hydrodynamical bounce for each of our rotating models.  
Solid lines are for the 11\,M$_\odot$ progenitor of Woosley \&
Weaver (1995).  Dashed lines are for model E15A from Heger et al.~(2000).
Specifications for the models are given in \S\ref{section:model} and Table \ref{tab:rot}.
Note that in none of the models shown here is a centrifugal barrier reached during collapse.  All models
undergo bounce near the nuclear phase transition in the inner core and have sub-Keplerian 
rotational velocities throughout the computational domain.
For clarity of presentation, the peak of the breakout pulse of electron neutrinos is not shown in the
upper-left panel.  The most rapid rotator ($P_0=1.25$ s) yields $L_{\nu_e}^{\rm peak}\simeq3.1\times10^{53}$
erg s$^{-1}$, whereas both E15A and the model with $P_0=2$ s have $L_{\nu_e}^{\rm peak}\simeq2.6\times10^{53}$
erg s$^{-1}$.  All other models have $L_{\nu_e}^{\rm peak}\simeq2.4\times10^{53}$
erg s$^{-1}$.  

As expected, we find that the higher 
the initial rotation rate, the lower the core temperatures of the PNS formed just after bounce.
On average, after electron neutrino 
breakout, this effect produces lower core $L_\nu$ and $\langle\varepsilon_\nu\rangle$ for shorter $P_0$.
This reflects the sensitivity of the neutrino emissivities to the local temperature.
The fractional differences in $\langle\varepsilon_\nu\rangle$ between the model with $P_0=1.25$ s and
a non-rotating model 200 ms after bounce are approximately 15\%, 17\%, and 30\% 
for $\nu_\mu$, $\bar{\nu}_e$, and $\nu_e$, respectively.  The same comparison for $L_\nu$ 
yields fractional differences of 75\%, 63\%, and 33\%
for $L_{\nu_\mu}$, $L_{\bar{\nu}_e}$, and $L_{\nu_e}$, respectively.
The difference in $L_{\nu_\mu}$ at 100\,ms after bounce between 
our slowest and fastest rotators is a factor of $\sim6$.  The neutrino spectral characteristics change
significantly and systematically as a function of $P_0$.
Despite the fact that model E15A has rapid rotation, it has a much larger $L_{\nu_e}$ after breakout than, for
example, the $P_0=2$ s 11 M$_\odot$ model, even though they have similar initial $\Omega(r)$ profiles
(see Fig.~\ref{plot:omegap2}).  This is due to the extended density profile of model E15A relative to the 11 M$_\odot$
model and the associated larger accretion luminosity.  

Figure \ref{plot:mdshbp} shows radial profiles of a number of quantities in a subset 
of the models of Fig.~\ref{plot:ltet} at a time-slice 105 ms after bounce.  The upper left panel shows the mass accretion rate
for the models with $P_0=1.25$ (dotted), 2 (dashed), 3 (dot-dashed), and 8 seconds (solid).  The shock is clearly visible in
each profile, the more rapid rotators having larger shock radii 105 ms after bounce.
Higher negative $\dot{M}$ implies larger PNS accretion rate and a larger accretion
luminosity.  This, coupled with the fact that the faster rotators achieve lower
core temperatures at bounce, explains  why the total luminosity is smaller for 
smaller $P_0$ in Fig.~\ref{plot:ltet}.  
The upper right panel shows $\Omega$ (positive numbers,
in units of $10^3$ rad s$^{-1}$) and $d\Omega/d\ln r$ (mostly negative numbers, in units of $10^3$ rad s$^{-1}$).
The same line styles are employed.  Note that the surface of the PNS (the neutrinosphere
at the peak of the $\nu_e$ spectrum) in this epoch
lies at approximately $50-70$ km (depending upon $P_0$).  Thus, if the mantle were
ejected at this instant, assuming strict angular momentum conservation, all models with $P_0\lesssim8$ s
form 10\,km neutron stars with $P<10$ ms.    In fact, the model with $P_0=8$ s would have a final spin period
of $\sim7$ ms.  The model with $P_0=2$ s would have a final angular velocity over $10^4$ rad s$^{-1}$.

The lower left panel of Fig.~\ref{plot:mdshbp}
shows the saturation magnetic field strength if the MRI operates everywhere, obtained simply
by equating the magnetic energy density with the azimuthal kinetic energy density (as in eq.~\ref{bsat}).
Figure \ref{plot:mdshbp} shows that  at radii near 100 km magnetic fields 
in the range $B_\phi\sim10^{14}-10^{15}$ G would be generic to relatively slowly rotating progenitors
(here typified by the $P_0=8$ s calculation, solid line).  For rapid rotators, much of the core has 
field strengths of $\sim10^{16}$ G and above.  
The results presented here for $B_{\rm Sat_\phi}$ are in rough agreement with the models of Akiyama et al.~(2003), 
but much higher than those presented in Fryer \& Warren (2004).  Comparing our Fig.~\ref{plot:mdshbp}
with their Fig.~13 we find that our field strengths are generally larger by $4-5$ orders of magnitude.

Although we have not included viscous dissipation in any of the models presented in this section,
in the lower right panel of Fig.~\ref{plot:mdshbp} we plot the magnitude of the viscous dissipation rate, assuming that the
MRI operates everywhere.  We combine eqs.~(\ref{dedt}) and (\ref{numri}) to obtain
\beq
\dot{q}_{\rm MRI}=\xi_{\rm MRI}\left(\frac{d\Omega}{d\ln r}\right)^2=\alpha H^2\Omega\left(\frac{d\Omega}{d\ln r}\right)^2,
\label{qmri}
\eeq
where we take $\alpha=0.1$.
The solid box shows a range of representative net neutrino energy deposition rates in the region exterior
to the PNS ($60<r<200$ km).  Typical net neutrino heating rates in the gain region  are 
$\dot{q}_\nu\sim100$ MeV nucleon$^{-1}$ s$^{-1}$$\sim9.6\times10^{19}$ erg g$^{-1}$ s$^{-1}$.
The box brackets the range $10^{18.5}\leq\dot{q}_{\rm MRI}\leq10^{20.5}$ erg g$^{-1}$ s$^{-1}$.
Of course, between the surface of the PNS and the shock the net heating rate varies
considerably, going from negative to positive at the gain radius $R_{\rm g}$.  This panel
is meant merely to underscore that a first comparison shows that the neutrino heating rate and the viscous
heating rate are comparable for rapid rotators.  If enough shear is present
so that the energy deposition rate implied by the lower right panel of Fig.~\ref{plot:mdshbp}
can be maintained, the dynamics of the collapsed core may be significantly affected  (\S\ref{section:noexp}).
This is particularly true for more rapid rotation because  the net neutrino heating rate in the gain region 
decreases as $\Omega$ increases and the viscous heating rate increases as $\Omega^3$.  
Thus, the relative importance of viscous heating with respect to neutrino heating is  compounded as 
the rotation rate increases.

For clarity of presentation we have not shown results for model E15A in Fig.~\ref{plot:mdshbp}.
The saturation magnetic field strength and viscous dissipation rate for this model are very similar
to the model with $P_0=2$ s in the lower left and right panels, respectively.
Model E15A has a larger negative $\dot{M}$ at $R_\nu$ and a larger $R_\nu$ than model $P_0=2$ s, but similar
shock radius.  Its shear and angular velocity profiles closely track those of model $P_0=2$ s for
$r\gtrsim 20$ km.  For $r\lesssim20$km model E15A has $\Omega>10^4$ rad s$^{-1}$, greater than the
corresponding number for model $P_0=1.25$ s.  These differences
between model E15A and model $P_0=2$ s reflect the difference in initial $\Omega(r)$ for
$r<600$ km (Fig.~\ref{plot:omegap2}) and the fact that the density profile for model E15A is 
much broader than that for $P_0=2$ s.

\subsection{Why Supernovae do not Explode in 1D}
\label{section:noexp}

Figure \ref{plot:mdshbp} shows that the viscous heating via the MRI can be comparable to the 
net neutrino heating in the gain region.  In \S\ref{section:dissipation}, we will show that
this effect is sufficient to yield explosions in some models.  In order to address the
question of why these models succeed, it is instructive first to consider why non-rotating
spherical models with the very best physics fail.  An elucidating and physical discussion
of the stalled shock dynamics can be found in Janka (2001).  We do not attempt to reproduce 
the detailed arguments presented there, but instead offer a semi-quantitative quasi-global criterion
for explosion in the spirit of the arguments presented in Thompson (2000).

First, we define a radial advective timescale across a pressure scale height $\tauadv=H/V_r$ and a timescale for net 
neutrino heating $\tauqdotnu=(P/\rho)/\dot{q}_\nu$, where $P$ is the thermal pressure, $\rho$ is
the mass density, $\dot{q}_\nu=H_\nu-C_\nu$, and $H_\nu$ and $C_\nu$ are the specific neutrino 
heating and cooling rates, respectively. Roughly speaking, steady-state accretion is maintained 
(explosions do not develop) in simulations of core-collapse in one dimension
because $\tauadv<\tauqdotnu$ in the gain region.  If this were not the case (that is, if $\tauadv>\tauqdotnu$),  
the thermal pressure  would increase between $R_{\rm g}$ and $R_{\rm sh}$ (or, not decrease as quickly
as material is advected into the cooling region) 
and the shock would move outward, finding a new equilibrium.  This condition, $\tauadv>\tauqdotnu$, 
is a necessary but not sufficient condition to guarantee explosion.  One must maintain the 
condition $\tauadv>\tauqdotnu$ for a time sufficient  to deposit roughly the binding energy of the
overlying mantle, $\sim10^{51}$ erg.   Thus, if the condition $\tauadv>\tauqdotnu$ 
is met for $t_i<t<t_f$ and 
\beq
E_{\rm binding}^{\rm mantle}<\int_{t_i}^{t_f}\int\dot{q}_\nu\rho\,\,d^3r\,dt,
\label{condition}
\eeq
then the mantle is unbound and an explosion results.  
The statement of eq.~(\ref{condition}) is equivalent to saying that the matter achieves
the "escape" temperature, obtained by equating the local specific thermal energy with the gravitational
potential energy (Burrows \& Goshy 1993; Burrows, Hayes, \& Fryxell 1995).

We illustrate these basic numbers in Fig.~\ref{plot:tcoolp2}, which shows $\tauqdotnu$ (solid line), $\tauadv$ (dashed line),
$\tau_{C_\nu}=(P/\rho)/C_\nu$ (dotted line), and $\tau_{H_\nu}=(P/\rho)/H_\nu$ (dot-dashed line) for a slowly rotating initial
progenitor ($P_0=8$ s, see Fig.~\ref{plot:ltet}) 130 ms after bounce.  The $\nu_e$ neutrinosphere ($R_{\nu_e}$) and the gain
radius ($R_{\rm g}$, where $\tau_{C_\nu}$ crosses $\tau_{H_\nu}$ and $\tauqdotnu$ diverges) are indicated by arrows 
at this snapshot in time.  Note that throughout the gain region ($r>R_{\rm g}$) $\tauadv<\tauqdotnu$.  The 
condition for explosion we suggest does not obtain.  Following the evolution in this model for another 400 ms, 
$\tauqdotnu/\tauadv$ is always greater than 1 at its minimum (which usually occurs very near, but exterior to,  $R_{\rm g}$).
We find very similar absolute values for the various timescales and ratios $\tauqdotnu/\tauadv$ for all 
models that fail to explode. Of course, one may
argue that the actual value of, for example, $\tauadv$, depends on its definition and that it may differ by order unity
from that presented in Fig.~\ref{plot:tcoolp2}.  For example, we define $\tauadv$ as $H/V_r$ rather than $r/V_r$ because
$\dot{q}_\nu$ depends fairly sensitively on radius near $R_{\rm g}$ since the cooling rates are highly dependent on
the local temperature.  Thus, the distance over which  $\dot{q}_\nu$ doubles near $R_{\rm g}$ is closer to $H$ than $r$.
This motivates our definition of $\tauadv$, but there is no reason {\it a priori} that $\tauadv$ should not differ from $H/V_r$
by, say, a factor of $\sim2$.    In addition, $\tauqdotnu$ as we have defined it corresponds to an $e$-folding time
for the local pressure, a relatively large change in the gain region.  It may be that well before a full $e$-folding
of $P$ is accomplished the system adjusts.  This is particularly true because the dynamical timescale
($\tau_{\rm Dyn}=H/c_s$ or $r/c_s$) is much shorter than all other timescales.  Thus, we cannot rule out
order unity corrections to $\tauqdotnu$.
For these reasons, that the ratio $\tauqdotnu/\tauadv$ is $\sim1.3$ at $r\sim125$ km
does not necessarily imply that if the 
net heating rate is increased by $\sim30$\%  the shock must begin to move outward.  
However, the ratio $\tauqdotnu/\tauadv$ is certainly indicative of the magnitude of the deficit in heating that
must be paid for explosion.  Remarkably, we show in \S\ref{section:dissipation} that the
criterion advocated here is quantitatively useful for assessing when explosions will develop.

Although simplistic,
the picture presented here, which corresponds in rough terms with the conditions derived in Janka (2001),
yields predictive power.  In {\it all} of our spherical, non-rotating models (see Thompson et al.~2003) 
$\tauadv$ is always less than $\tauqdotnu$ in the gain region over the full post-bounce evolution.  
Similarly, in model E15A and the models with $P_0=2$, 3, 4, 5, and 8 seconds, despite the wide range of
mass accretion rates and post-bounce structures, the shock stalls in each model for the duration of the calculation
and $\tauadv/\tauqdotnu$ is always less  than unity (when dissipation is not included).\footnote{The model with $P_0=1.25$ s
never develops a well-defined gain region and the shock never stalls completely.
The centrifugal support (the decrease in the gravitational potential) as the shock is
launched is sufficient to keep the shock moving, even though Keplerian rotational velocities are
not obtained anywhere in the calculation.  Matter velocities never exceed $2\times10^8$ cm s$^{-1}$ during 
the calculation and the internal energy of the ejected matter is small because there is little, if any, neutrino 
heating as the explosion develops. Roughly estimating the asymptotic energy, we obtain $<10^{50}$ erg.}
Exceptions to this rule occur only during very dynamic
episodes. For example, the instant the Si shell encounters
the shock,  the density decreases, $\dot{M}$ decreases, and
the shock moves out.  In the instant between the two epochs of quasi-steady-state accretion,
$\tauadv$ can be greater than $\tauqdotnu$, but this is a  transient and although the
shock moves outward in radius, no explosion is obtained.  In marked contrast, if one injects energy into the
gain region, say via viscous dissipation with rate $\dot{q}_{\rm MRI}$, then the relevant comparison is between
$\tauadv$ and $\tauqdottot =(P/\rho)/\dot{q}_{\rm TOT}$, where $\dot{q}_{\rm TOT}=H_\nu-C_\nu+\dot{q}_{\rm MRI}$.
Adding $\dot{q}_{\rm MRI}$ to $\dot{q}_\nu$ decreases the net heating timescale from $\tauqdotnu$ to $\tauqdottot$.
If $\dot{q}_{\rm MRI}$ is large enough to yield $\tauqdottot<\tauadv$ and condition eq.~(\ref{condition})
is met, an explosion results. We show in \S\ref{section:dissipation} that this mechanism succeeds.

\section{Comparing Viscous Processes}
\label{section:compare}

In this section we compare quantitatively the magnitude of the neutrino viscosity with the
turbulent viscosity of the MRI in the dense protoneutron star core where neutrinos are diffusive.
We also compare the linear growth rate for purely hydrodynamical convection and the MRI
in the semi-transparent gain region with the neutrino damping rate.  
There have been a number of papers concerning the possibility that 
convection persists in PNS interiors during the accretion/explosion 
phase and afterwards, during Kelvin-Helmholtz cooling (Mayle 1985; Mayle \& Wilson 1988; Wilson \& Mayle 1988;
Burrows 1987; Burrows \& Lattimer 1988; Thompson \& Duncan 1993; Keil \& Janka 1995; Pons et al.~1999;
Thompson \& Murray 2001).  
If the neutrino viscosity ($\xi_\nu$) is comparable to or larger than the viscosity of the MRI ($\xi_{\rm MRI}$) or convection, 
any turbulent fluctuations generated by these instabilities may be damped.  Similarly, 
convective or magneto-convective modes
in the gain region may also be damped if the neutrino damping rate ($\Gamma_{\nu}$) is larger than the
linear growth rates for convection ($\sqrt{|N^2|}$) or the MRI ($\Gamma_{\rm MRI}\sim\Omega$).

In Fig.~\ref{plot:nunu2}  we plot $\xi_\nu$, $\xi_{\rm MRI}$, $\Gamma_{\nu}$, $\Gamma_{{\rm MRI}}$, and $\sqrt{|N^2|}$
for the model with $P_0=2$ s at a time 105 ms after bounce, as in Fig.~\ref{plot:mdshbp}.
The left panel shows $\xi_\nu$ for each neutrino species and $\xi_{\rm MRI}$ in the inner region 
of the PNS, inside the neutrinospheres  for $\nu_e$ and $\bar{\nu}_e$ neutrinos  at their respective spectral peaks.  
In this diffusive regime $\langle \lambda_\nu\rangle$ is less than or much less than $H$ and
a comparison between the neutrino viscosity and the turbulent viscosity of the MRI is appropriate.
For the purposes of comparison we have taken $\xi_{\rm MRI}=\alpha H^2\Omega$ with $\alpha=0.1$.   
The viscosity caused by the MRI clearly dominates neutrino viscosity by $2-3$ orders of magnitude
over the entire profile. Thus, within the PNS, when the condition for instability to the MRI is met (eq.~\ref{gso}),
the MRI will operate and dominate the viscosity.  Even for the slowest rotator considered here ($P_0=8$ s),
$\xi_{\rm MRI}$ is greater than $\xi_\nu$ by an order of magnitude or more throughout the PNS interior.

In the right panel of Fig.~\ref{plot:nunu2} we plot $\Gamma_{\nu}$  for each neutrino
species and $\Gamma_{{\rm MRI}}$ as a function of radius in the semi-transparent region from roughly
the surface of the PNS out to the shock radius.   Again, the linear growth rate of the
MRI dominates the neutrino damping rates.  At $r\sim50$ km the ratio $\Gamma_{{\rm MRI}}/\Gamma_{\nu}\simeq10$.
In the gain region ($r\sim150$ km), where the comparison is most important (that is where condition eq.~\ref{gso}
is always met)  $\Gamma_{{\rm MRI}}/\Gamma_{\nu}\simeq10^3$.
For the model with $P_0=8$ s,  $\Gamma_{{\rm MRI}}/\Gamma_{\nu}\sim100$ in the gain region.
For comparison, we also show the linear growth rate for convection ($\sqrt{|N^2|}$,
dot-dashed line).  As $P_0$ decreases in our progenitor, $\sqrt{|N^2|}$ in the gain region stays roughly constant while
$\Gamma_{{\rm MRI}}$ decreases.  In the model with $P_0=8$ s, $\sqrt{|N^2|}/\Gamma_{{\rm MRI}}\sim10$ 
at $t\sim100$ ms after bounce, at $r\sim150$ km.
Importantly, both $\Gamma_{{\rm MRI}}$ and $\sqrt{|N^2|}$ dominate $\Gamma_{\nu}$
and we therefore expect well-developed, high Reynolds number convection/magneto-convection
throughout the gain region in multi-dimensional models.

Although the profiles evolve considerably in time, in general, 
we find that much of the PNS core and all of the gain region is unstable to
the MRI.  In the cooling region just above the PNS ($50\,\,{\rm km}\lesssim r\lesssim100\,\,{\rm km}$), 
$N^2$ in eq.~(\ref{gso}) is sufficiently large and positive to counter much of the negative shear profile
and the MRI is stabilized. However, we expect that a multi-dimensional simulation of the
MRI in this context may show magneto-convective plumes and filaments penetrating the
cooling region, as in the purely hydrodynamical calculations of Burrows, Hayes, \& Fryxell (1995).
Deeper in the PNS, beneath the $\bar{\nu}_e$ neutrinosphere,  the profile is unstable.
The region $20\,\,{\rm km}\lesssim r\lesssim50\,\,{\rm km}$ is unstable to the MRI throughout most of 
the post-bounce epoch in the $P_0=2$ s model.  For slower initial rotation, this region of instability 
shrinks in radial extent.
Imposing $\Omega(r)$ as in eq.~(\ref{omegaprofile}), we find that the
very deep interior ($r\lesssim10$ km) has positive or near-zero shear (see upper right panel
of Fig.~\ref{plot:mdshbp}) and is stable to the MRI.

The region of very strong negative shear between $10\lesssim r\lesssim20$ km in the upper right
panel of Fig.~\ref{plot:mdshbp} contains strong stabilizing entropy gradients and is stable over
most of the evolution in all of the $P_0$ models considered.  Importantly, even though 
eq.~(\ref{gso}) indicates that this region is stable to the MRI and convection, it is still possible to 
transport angular momentum and dissipate shear energy,
albeit on a modified timescale (Spruit 2002).  In stably stratified regions with differential  
rotation and a thermal diffusivity, $\kappa$, Spruit (2002) argues that  magnetic interchange 
instabilities lead to an effective viscosity of (see his eq.~32)
\beq
\xi_{\rm S}\sim H^2\Omega\left(\frac{\Omega}{N}\right)^{1/2}\left(\frac{\kappa}{r^2N}\right)^{1/2},
\label{nuspruit}
\eeq
where we have assumed a correlation length of $H$ instead of $r$.
Estimating $\xi_{\rm S}$ with $\kappa=c\langle\lambda_\nu\rangle/3$ and comparing with the left panel of Fig.~\ref{plot:nunu2},
we find that in the region $10\lesssim r\lesssim20$ km $\xi_{\rm S}$ is less than $\xi_{\rm MRI}$ by approximately 
one order of magnitude and greater than $\xi_\nu$ by  approximately two orders of magnitude.  
This  implies a viscous timescale  of $\tau_{\xi_{\rm S}}=r^2/\xi_{\rm S}\sim1-5$ seconds during
the post-bounce epoch considered here.  Interestingly,   $\tau_{\xi_{\rm S}}$ may be different (possibly longer) than 
the Kelvin-Helmholtz timescale for late-time cooling.

Although the comparison of Fig.~\ref{plot:nunu2} shows that the two parts of the
generalized Solberg-H\o iland criterion are of similar magnitude, it is difficult to
estimate the relative importance of convection versus the MRI in the non-linear regime without 
recourse to multi-dimensional simulations.  Taking our estimates of the turbulent viscosity for
each process (eqs.~\ref{numri} and \ref{nucon}) we have 
\beq
\frac{\xi_{\rm Con}}{\xi_{\rm MRI}}=
\frac{1}{3\alpha}\left(\frac{V_{\rm Con}}{H\Omega}\right).
\eeq
Taking $V_{\rm Con}\sim5\times10^8$\,cm\,s$^{-1}$ 
as representative of the results from multi-dimensional
simulations and $H\Omega\sim10^9$ cm s$^{-1}$ ($P_0\lesssim3$ s) we see that for $\alpha=0.1$,
$\xi_{\rm Con}$ and  $\xi_{\rm MRI}$ are of the same order of magnitude.  
We reiterate, however, that even the sign of the convective stress is uncertain (\S\ref{section:convection})
so the comparison presented here is quite simplistic.

\section{Core Collapse with Rotation \& Viscous Dissipation}
\label{section:dissipation}

We now provide results from a subset of the rotating models considered in
\S\ref{section:rotation}, including viscous dissipation and angular momentum transport.

The model with $P_0=2$ seconds (see \S\ref{section:rotation} and
Figs.~\ref{plot:omegap2}, \ref{plot:ltet}, \ref{plot:mdshbp}, and \ref{plot:nunu2}) serves
to illustrate the effects of viscous dissipation most clearly.  Without including dissipation, 
at 290 ms after bounce there is a well-defined gain region with $R_{\rm g}\simeq 115$ km 
and a shock radius of $\sim275$ km. The model is not exploding.  It has timescale profiles 
very much like those in Fig.~\ref{plot:tcoolp2}  for the model with $P_0=8$ s with $\tauadv<\tauqdotnu$ in the gain region
(see \S\ref{section:noexp}).  In this late post-bounce epoch we restart the calculation including viscous 
dissipation, $\dot{q}_{\rm MRI}$ with $\alpha=0.1$ (see eq.~\ref{qmri}).   The reason for restarting the
simulation with dissipation nearly 300 ms after bounce instead of just 100 or 200 ms after bounce
is that by this time in this progenitor the Si shell has fallen through the shock and the dynamical
phase of shock expansion associated with this event is over.  Because the system has settled into
a steady-state accretion flow it is more easily diagnosed as we add heating by viscous dissipation.
For this model, we include $\dot{q}_{\rm MRI}$ {\it only} in the
gain region above the PNS surface for $r>120$ km.  Figure \ref{plot:sinkmp} shows the relative
contributions to $\dot{q}_{\rm TOT}=\dot{q}_\nu+\dot{q}_{\rm MRI}$ (solid line) at a time 340 ms after bounce.
The heating rate from viscous dissipation ($\dot{q}_{\rm MRI}$), the total
neutrino heating rate ($\dot{q}_\nu$), and the contributions to $\dot{q}_\nu$
from $\nu_e$ and $\bar{\nu}_e$ neutrinos are shown.  

Complementary to Fig.~\ref{plot:sinkmp}, but more germane to the issue of explosion, is Fig.~\ref{plot:tcoolp},
in which we show all of the timescales presented in Fig.~\ref{plot:tcoolp2} and
discussed in \S\ref{section:noexp}, but include $\tau_{\dot{q}_{\rm MRI}}=(P/\rho)/\dot{q}_{\rm MRI}$
(short dashed line).  We distinguish between $\tau_{\dot{q}_\nu}=(P/\rho)/\dot{q}_\nu$ (dot-long dashed line) and 
$\tau_{\dot{q}_{\rm TOT}}=(P/\rho)/(\dot{q}_\nu+\dot{q}_{\rm MRI})$ (solid line), where the former contains
contributions only from neutrino heating and the latter includes both neutrino heating and
viscous dissipation of shear energy.  As shown clearly in Fig.~\ref{plot:tcoolp}, viscous dissipation 
is sufficient to make  $\tau_{\dot{q}_{\rm TOT}}$  less than $\tau_{\rm Adv}$.  As per the discussion
in \S\ref{section:noexp} this is suggestive of a condition for a thermally-driven explosion.
In this simulation and others, we find that if the condition  $\tau_{\rm Adv}/\tau_{\dot{q}_{\rm TOT}}>1$ is met just outside
$R_{\rm g}$, the pressure increase is communicated to the rest of the gain region on a dynamical
timescale and the shock expands.   That the number on the right-hand-side of this inequality is
so close to unity is surprising considering the ambiguities in defining $\tau_{\rm Adv}$ and $\tau_{\dot{q}_{\rm TOT}}$.
Nevertheless, we find that it is accurate.  The region in Fig.~\ref{plot:tcoolp} 
where $\tau_{\rm Adv}/\tau_{\dot{q}_{\rm TOT}}>1$ grows in time.  
The gain region expands as the shock moves outward. When the thermal energy of the gas
exceeds the gravitational binding energy (and the escape temperature obtains) an explosion develops (eq.~\ref{condition}).

The explosion is not driven by viscous dissipation alone.  
It is triggered by the combined action of $\dot{q}_{\rm MRI}$ and $\dot{q}_\nu$.  In fact, just $\sim35$ milliseconds
after the condition $\tau_{\rm Adv}/\tau_{\dot{q}_{\rm TOT}}>1$ is met in Fig.~\ref{plot:tcoolp},
$\tauadv$ has increased enough so that  $\tauqdotnu<\tauadv$.   That is, neutrino heating
alone is then sufficient to drive the explosion.  In these $\sim35$ milliseconds $\dot{q}_{\rm MRI}$ 
deposits only $\sim2.1\times10^{49}$ erg in the gain region.  The neutrinos account for nearly five times that 
(Fig.~\ref{plot:sinkmp}).
This illustrates explicitly that any heating source that contributes an additional $\sim20-50$\%
to the net neutrino heating rate should be sufficient in most models to instigate explosion (see \S\ref{section:noexp}
and Fig.~\ref{plot:tcoolp2}).  In the model described here it takes $\sim120$ ms after the condition
$\tau_{\rm Adv}/\tau_{\dot{q}_{\rm TOT}}>1$ is met for matter in the gain region to attain positive velocity.
In this time period, $\sim3.1\times10^{49}$ erg of shear energy is dissipated, whereas $\sim1.7\times10^{50}$
erg is deposited by neutrinos.  Because the explosion is initiated at late times
(we did not add $\dot{q}_{\rm MRI}$ until $\sim290$ ms after bounce), the extrapolated asymptotic
energy of this explosion is rather small.  At the end of the calculation, $\sim1.1$ seconds 
after bounce, taking rough account of the kinetic, thermal, and gravitational potential energies we estimate
an asymptotic kinetic energy of $\sim1.6\times10^{50}$ erg.  Including the energy gained by recombination to nuclei of
the $\sim0.03$ M$_\odot$ with positive velocities at the end of the calculation adds $\sim5.1\times10^{50}$ erg,
or a total asymptotic energy of $\sim6.7\times10^{50}$ erg.  Continued energy injection by neutrino heating
will increase this estimate.  The protoneutron star gravitational mass at the end of this calculation is $\sim1.404$ M$_\odot$.

Including viscous dissipation only in the region $r>120$ km is quite conservative.  In fact, although there
are strong stabilizing entropy gradients in the cooling region, because of the strong negative shear there,
part of the cooling layer can be driven unstable by the MRI.  In addition, because this region does not 
encompass many pressure scale heights, we expect magneto-convective plumes to overshoot the gain region
and enter the neutrino cooling layer.  For these reasons we carried out two calculations analogous to the
one just described, but which include viscous heating in the cooling region (for $r>60$ km and for $r>80$ km).
Both models explode at earlier times and with higher neutrino luminosities than the model with dissipation exterior to $r>120$ km.
Because the explosion is instigated earlier, these models leave behind neutron stars of smaller mass and yield explosions of higher energy.


Similar numerical experiments may be carried out with different initial rotation periods and
$\alpha$s.  For larger $\alpha$ and given $P_0$, the viscous time is shorter, shear energy is dissipated 
more rapidly (though roughly the same total energy is dissipated), the condition that 
$\tauadv/\tau_{\dot{q}_{\rm TOT}}>1$ is more easily obtained, and explosions take less time to develop.  
For smaller $\alpha$, the threshold set by the condition $\tauadv/\tau_{\dot{q}_{\rm TOT}}>1$
may not be reached or it may happen on a timescale too long to explore in
the models presented here.\footnote{Due to numerical limitations, including shock resolution and
Courant-limited time stepping, it is difficult to push our calculations much beyond 1 s after bounce.}
For example, taking $\alpha=0.01$, $\dot{q}_{\rm MRI}/\dot{q}_{\nu}\sim0.03$
in the gain region for the $P_0=2$ s model just considered (see Fig.~\ref{plot:sinkmp}),  $\tauadv/\tau_{\dot{q}_{\rm TOT}}<1$ 
throughout the gain region, and no explosion is obtained.

For shorter $P_0$ (faster rotation), more shear energy is available in the gain region and at a given  $\alpha$
an explosion develops more quickly.   In fact, the viscous heating rate  at a given radius 
scales with $\Omega^3$ and, all else being equal, if $P_0$ is decreased from $2$ s to $4$ s,
the heating rate in the gain region decreases by a factor of $\sim6$.  
Thus, for a given $\alpha$, beyond a threshold $P_0$  the difference between
$\tauadv$ and $\tau_{\dot{q}_{\nu}}$ cannot be compensated for by including $\dot{q}_{\rm MRI}$
and no explosion obtains.  However, this threshold depends on the extent of the region of viscous heating.

It is perhaps artificial to begin including
viscous dissipation so long after bounce.  The purpose of the previous experiment was merely to 
clearly isolate the physics associated with shear energy dissipation.
In reality, the magnetic field
grows from an arbitrarily small seed field exponentially on a timescale $\sim\Omega^{-1}$.
We expect a saturation magnetic field to obtain and fully developed magneto-convection to appear
just $10-100$ ms after bounce in the region exterior to the PNS.  For this reason, 
we have run a number of models including
$\dot{q}_{\rm MRI}$ after bounce, following the full evolution.

To this end, we conduct two tests with the model E15A (see \S\ref{section:progenitor}, 
Table \ref{tab:rot}, and Figs.~\ref{plot:omegap2} and \ref{plot:ltet}).
In one case, we include viscous dissipation at all radii.  In the second calculation
we turn on viscous dissipation only when the generalized Solberg-H\o iland criterion 
indicates that the region is unstable (eq.~\ref{gso}).  This latter case is probably more realistic within the
context of the spherical models presented in this paper.   Both models employ $\alpha=0.1$ and
do not transport angular momentum until after bounce.  In both cases, we obtain explosions that develop on
the relatively short timescale of $\sim50-100$ ms after bounce.  

The explosions in this model
are qualitatively different from the explosion obtained by waiting 
until a well-defined gain region forms and then adding viscous dissipation 
(Figs.~\ref{plot:sinkmp} and \ref{plot:tcoolp}). In non-exploding models, the bounce
shock stalls immediately after formation, but then moves slowly outward in radius as the
cooling and gain regions are formed until the shock reaches $R_{\rm sh}\sim200$ km (about
$\sim100-200$ ms after bounce).  Its outward
progress is halted primarily by neutrino losses in the forming cooling region.  Usually,
the shock radius recedes slowly after reaching its maximum radius and sits for many 
hundreds of milliseconds at $r\sim150$ km.
Indeed, this evolutionary sequence is borne out in model E15A without viscous dissipation;
no explosion occurs.
Including viscous heating in E15A, the evolution is very different.  
Neutrino losses behind the shock (as $R_{\rm sh}$ moves from $\sim$80 km to $\sim$200 km)  
are partially compensated by viscous heating.  This extra energy deposition allows the shock to
continue its outward progress, putting matter higher in the PNS gravitational well, and
explosions are obtained.
Because the shear in this model is rather large, the region 
behind the shock is unstable to the MRI and viscous dissipation operates.
Stabilizing entropy gradients are insufficient to make $N^2+d\Omega^2/d\ln r$ positive (eq.~\ref{gso}).
Because both the calculation  with dissipation everywhere and  the calculation respecting eq.~(\ref{gso})
have viscous heating that partially compensates neutrino cooling as the shock is moving outward, the 
same qualitative evolution is obtained in both models.
One important difference between them is that,
on average, the model with dissipation throughout the profile has more heating behind the shock
than the model that employs the generalized Solberg-H\o land criterion.  This means that the explosion develops
faster and leaves behind a less massive neutron star.  Although we are able to follow only
the first $\sim200$ ms after explosion ($\sim500$ ms after bounce), the difference between the PNS mass at the end of our
simulations is roughly $\sim0.15$ M$_\odot$, the larger baryonic mass being $\sim1.4$ M$_\odot$.

Figure \ref{plot:angp} shows the angular momentum evolution for the E15A model with dissipation at all radii
as a function of the enclosed mass.  For reference, the dotted line shows $j_{\rm Kep}=r^2\Omega_{\rm Kep}$
at bounce ($t=0$).  
The profile $j(M)$ is shown at bounce and 22, 100, 
200, and 350 ms after bounce.   The transport of angular momentum is clear. 
For high viscosity, $\Omega$ can decrease appreciably throughout the PNS core during explosion,
providing a mechanism for at least partial spindown of the young PNS.
Figure \ref{plot:angp} can be compared with
the three-dimensional simulations of Fryer \& Warren (2004), who found significant angular momentum
transport in the deep core of their rotating progenitors (see their Figs.~9 and 10).  At $M\sim0.5$ M$_\odot$
they find that $j$ decreases by a factor of $\sim4$ in $\sim 200$ ms.  
We find a similar magnitude 
change in $j$ at this $M$, implying that the magnitude of the intrinsic angular momentum transport
in the simulations of Fryer \& Warren (2004) is comparable to the transport caused here by our explicit viscosity.

Similar to the E15A models just described, we computed the evolution of
$P_0=1.25$, 2, 3, 4, and 6 s models with the 11 M$_\odot$ progenitor including viscous dissipation with $\alpha=0.1$.
Like the E15A models,
we calculate a set of models with dissipation everywhere at all radii and a set of models
respecting the generalized Solberg-H\o iland stability criterion.  All models with $P_0<4$ s
explode energetically before a well-defined gain region is established. The model with $P_0=4$ and dissipation everywhere also explodes,
but on a longer timescale ($\sim200$ ms after bounce).   The model with $P_0=6$ s does not explode.
The $P_0=4$ s model with dissipation only in regions where the
MRI is unstable also does not explode.  In this model, during the post-bounce epoch, $\dot{q}_{\rm MRI}$
was not large enough to force $\tau_{\dot{q}_{\rm TOT}}<\tauadv$ in the gain region.  
In addition, just after bounce the shear is not strong enough to drive a large portion of the neutrino cooling
region unstable to the MRI.  For this reason, explosions qualitatively similar to the E15A 
models do not obtain.  In summary, our calculations and estimates of the energy deposition rate
of viscous dissipation show that for initial spin periods in the range $P_0\lesssim4-6$ s in our 11 M$_\odot$
progenitor, and employing the angular momentum profile dictated by equation (\ref{omegaprofile}),
energetic supernova explosions can be obtained for $\alpha\simeq0.1$ with dissipation everywhere.  
We therefore quote an initial spin period required to significantly modify the dynamics and outcome 
of one-dimensional simulations of core-collapse supernovae of $P_0\sim5$ s.
More detailed predictions must await multi-dimensional MHD simulations.


Typical results for the explosions we obtain are shown in Figs.~\ref{plot:vmrp2} and \ref{plot:masstee}. 
The former shows  snapshots of the velocity profile at various times during the development of explosion in our model with $P_0=2$ s.
This particular model included viscous dissipation at all radii, but the results for this initial spin period are very much like those with
dissipation only in regions where the generalized Solberg-H\o iland stability criterion indicates instability.
The shock stalls and an accretion phase begins, but this model never reaches a steady state with a well-defined gain region.
The constant injection of thermal energy by viscous dissipation keeps the pressure behind the
shock relatively high and it moves out slowly, putting post-shock material higher in the PNS's
potential well.  An explosion develops on a timescale of just $\sim100$ ms after bounce.
In Figure  \ref{plot:masstee}  we show the full history
of the mass element trajectories in radius as a function of time in the model with $P_0=4$ s and dissipation everywhere.  
This explosion takes considerably more time to develop.  In this model it is primarily the viscous heating in the cooling
region that instigates explosion after a well-defined gain region has developed.  
The heavy solid lines show the trajectories of the 1.0, 1.2, 1.3, and 1.4 M$_\odot$ mass elements.
The simulation is very well resolved; only every tenth mass zone is shown.  
The neutron star mass at the end of the calculation is $\sim1.35$ M$_\odot$.

As an aside, 
in Fig.~\ref{plot:csrp} we provide two snapshots of the velocity profile (solid lines) in a representative exploding model.
For comparison, we also include the sound speed profile.  The two snapshots are separated by $\sim$100 ms
in time and the shock moves from approximately 750 km to 2000 km.  The large dot at $\sim$850 km
marks the sonic point and the emergence of a transonic PNS wind (Duncan et al.~1986; Woosley et al.~1994; 
Takahashi, Witti, \& Janka 1994; Burrows, Hayes, \& Fryxell 1995).  
Profiles of entropy, mass loss rate, and electron fraction
all show that a quasi-steady-state solution, quantitatively similar to that expected from the models
of Thompson et al.~(2001) and the analytic estimates of Qian \& Woosley (1996) is obtained as the PNS
begins the Kelvin-Helmholtz cooling epoch.
In all of our successful exploding models, a transonic wind is formed.
At the end of this calculation ($P_0=2$ s, $\alpha=0.1$), 100 ms after the second snapshot in 
Fig.~\ref{plot:csrp}, the mechanical power of the wind is $\gtrsim10^{50}$ erg s$^{-1}$.  
If strong transonic PNS winds are a generic feature of the core-collapse supernova
phenomenon, as implied by our calculations, this conclusion may have implications for 
late-time fallback and the mass cut (Woosley \& Weaver 1995).

\section{Summary, Conclusions, \& Implications}
\label{section:summary}

In this paper we forward the idea that some of the gravitational binding energy of
core collapse may be trapped in differential rotation and that this free energy can be tapped
on a viscous timescale ($\sim1$ s in the gain region) and deposited as thermal energy in the gas. 
We identify magnetic stresses generated by the magnetorotational instability
as the most likely origin of this viscous energy dissipation. 
We have constructed models of core collapse, adding rotation and viscosity
to our existing algorithm for spectral, multi-angle radiation hydrodynamics.
We find that viscous dissipation can enhance the heating in the gain region 
sufficiently to yield explosions in models that would otherwise fail.
Initial spin periods of the roughly solid-body iron core should be $P_0\lesssim5$ seconds
for this mechanism to be robust.  

We emphasize several important points about this result.
First, our simulations are one-dimensional and as such do not include the convective
motions that have been shown to at least aid, if not enable, explosions (Herant et al.~1994; 
Burrows, Hayes, \& Fryxell 1995; Janka \& M\"{u}ller 1996; Fryer et al.~1999; Janka et al.~2002; Buras et al.~2003).   We expect
that multi-dimensional models with {\it slower} initial progenitor rotation periods, including
the effects of viscous dissipation consistently, will also yield explosions.  
Second, the effects of dissipation depend on where, when, and how quickly dissipation acts.  
Improved understanding of these issues will require MHD simulations and further progenitor modeling.
Third,  the values of  $P_0$ we require to significantly effect the dynamics of core collapse 
are fully consistent with the rotating progenitor models of Heger et al.~(2000).  However, there is considerable 
uncertainly in these progenitor rotation rates.  In particular, the models of Heger et al.~(2003) and Heger et al.~(2004),
which attempt to include angular momentum transport by magnetic torques consistently in one dimension, find 
much slower rotation rates for the iron core just prior to collapse.   For low mass progenitors in the range $12-15$ M$_\odot$,
the rotation periods obtained by  Heger et al.~(2004) are larger than our $P_0\lesssim5$ s fiducial bound.  
Heger et al.~(2004) do,
however, find that the rotation period decreases with increasing
progenitor mass.  Thus, without a detailed study beyond the scope of this
paper, we cannot make a definite statement as to whether rotational
effects are important in their core-collapse progenitor models.

Finally, our limit on $P_0$ is, at face value, inconsistent with the observed distribution of spin periods
of young pulsars, which are in the range of many tens to many hundreds of milliseconds (Kaspi \& Helfand 2002).  
There is, of course, a danger in such comparisons since it is very difficult to establish the
true birth periods of neutron stars observationally.  It is also unclear if the spindown of very young neutron
stars is well-understood theoretically as there are a number of ways to remove 
angular momentum from the neutron star during the explosion phase (as in the models presented here
with viscous dissipation) 
or during the Kelvin-Helmholtz 
cooling epoch (\S\ref{section:motivate}).  Given these uncertainties, we reiterate that the range of rotation
rates considered here are expected theoretically in models of magnetar formation
(Duncan \& Thompson 1992; Thompson \& Duncan 1993; Kouveliotou et al.~1999), a possibly
large subset of all supernovae.

Our work has highlighted two primary impacts of rotation on the mechanism of core-collapse
supernovae: 1) Most importantly, rotational kinetic energy tapped by viscous processes can
help drive explosions (\S\ref{section:dissipation}), and 2) Rotation lowers the effective gravity in the
core, increasing the radius of the stalled shock and the size of the gain
region (\S\ref{section:rotation}, Fig.~\ref{plot:mdshbp}).  Since ejection
is inhibited by the deep potential well, this helps facilitate explosion.
In addition to these, there are several multi-dimensional effects of
rotation on core-collapse that remain to be fully explored (Burrows, Ott, \& Meakin 2003):  1) Rotation
might generate vortices that can dredge up heat from below the
neutrinospheres and thereby enhance the driving neutrino luminosities.
2) Rotation can result in a large pole-to-equator anisotropy in the mass
accretion rate after bounce due to the centrifugal barrier along the
poles. Finally, 3) rotation results in pole-to-equator anisotropies in the
driving neutrino flux that have never before been accurately calculated.
Since the polar neutrino flux should be enhanced and the polar mass flux
should be lowered (point 2), explosion along the poles with bipolar
morphology seems that much more likely (Kotake, Yamada, \& Sato 2003;
Burrows \& Goshy 1993; Burrows, Ott, \& Meakin 2003).  Thus, such bipolarity, and the consequent optical
polarization of the debris, are not exclusive signatures of MHD-driven
explosions (Akiyama et al. 2003; Symbalisty 1984), but may be a natural
consequence of any rapidly rotating models.  

In addition to the effects of rotation on the supernova mechanism, we have also shown how the
observational signature of core collapse might be affected by rotation by identifying the
systematics in the neutrino luminosity and average energy as a function of time,
initial rotation rate, and progenitor (\S\ref{section:rotation}, Fig.~\ref{plot:ltet}).
If the neutrino radiation from the next galactic supernova is observed in modern neutrino
detectors like SuperKamiokande and SNO, we may hope to distinguish rapidly rotating core collapse
via its neutrino signature.  In addition to the difference highlighted in Fig.~\ref{plot:ltet}
we note that rapidly rotating models may leave an appreciable amount of shear energy ($\sim10^{52}$ erg)
trapped in the inner core, which is stable to the MRI and convection.  The timescale to dissipate
this energy is uncertain, but could be $\sim 1-10$ seconds or longer (eq.~\ref{nuspruit}; Spruit 2002) as the PNS cools.  The
timescale for dissipation of this energy will likely differ from the Kelvin-Helmholtz timescale (Burrows \& Lattimer 1986;
Pons et al.~1999).  
Thus, rapidly rotating PNSs may have a unique long-timescale neutrino signature.

Finally, we have assessed the relative role of neutrino viscosity, the turbulent viscosity caused by
the magnetorotational instability, and the viscosity of purely hydrodynamic convection.
We find that, for the rotation rates considered here, the MRI and convection may be comparable in
importance, but that both dominate microscopic neutrino viscosity
everywhere and at all times during the collapse and accretion epochs.  Turbulent
magneto-convection should be generic for all modestly rotating progenitors and of
critical importance to the dynamics of models with fairly rapid rotation, $P_0\lesssim5$ s
(Duncan \& Thompson 1992; Thompson \& Duncan 1993).
Thus, in models with these rotation rates, magnetohydrodynamic effects 
should not be ignored. 
The simplicity of the algorithm employed here, which assumes spherical symmetry in
a system that so clearly violates it and an $\alpha$ viscosity, makes it clear that more
work must be done to ascertain the full effects of MHD in rotating core collapse.

\acknowledgments

We gratefully acknowledge helpful conversations with Jon Arons, Anatoly Spitkovsy, and Yoram Lithwick and
we extend our gratitude to Craig Wheeler and David Meier for a careful reading of the text.
We thank Frank Timmes for making the Helmholtz EOS available.
We also thank Alex Heger for providing rotating pre-collapse progenitors.
T.~A.~T. is supported by NASA through Hubble Fellowship
grant \#HST-HF-01157.01-A awarded by the Space Telescope Science
Institute, which is operated by the Association of Universities for 
Research in Astronomy, Inc., for NASA, under contract NAS 5-26555.
E.~Q. is supported in part by NSF grant AST 0206006, NASA grant NAG5-12043,
an Alfred P. Sloan Fellowship, the David and Lucile Packard Foundation, and a Hellman Faculty Fund Award.
A.~B. is supported by the Scientific Discovery through Advanced 
Computing (SciDAC) program of the DOE, grant number DE-FC02-01ER41184.


\newpage
\begin{table}
\begin{center}
\caption{Rotating Core-Collapse Without Dissipation \label{tab:rot}}

\begin{tabular}{lcccccc}
\\
\hline \hline
\multicolumn{1}{c}{} &\multicolumn{1}{c}{} &
\multicolumn{1}{c}{} &\multicolumn{1}{c}{}&\multicolumn{1}{c}{}&\multicolumn{1}{c}{} \\

\multicolumn{1}{c}{Name} &
\multicolumn{1}{c}{$\beta^{\rm Rot}_i$ }&
\multicolumn{1}{c}{$\beta^{\rm Rot}_b$ }&
\multicolumn{1}{c}{$\beta^{\rm Rot}_{500}$ }&
\multicolumn{1}{c}{$E^{\rm shear}_b$ ($10^{51}$ erg)}&
\multicolumn{1}{c}{$E^{\rm shear}_{500}$ ($10^{51}$ erg)}\\

\\
\hline

& & & \\ 

$P_0=1.25$ s & 0.0035 & 0.059 & 0.070 & 21.8 & 32.9  \\

& &  &  \\ 

$P_0=2.0$ s & 0.0014 & 0.028 & 0.039 & 11.4 & 25.0     \\

& &  & \\ 

$P_0=3.0$ s & $6.0\times10^{-4}$ & 0.012 & 0.020 & 4.98 & 13.9 \\

& &  &  \\ 

$P_0=4.0$ s & $3.4\times10^{-4}$ & 0.0065 & 0.012 &  2.46 & 8.39 \\

& &  & \\ 

$P_0=5.0$ s & $2.2\times10^{-4}$ & 0.0042 & 0.0077 & 1.61 & 5.56 \\

& &  &  \\ 

$P_0=8.0$ s &  $8.5\times10^{-5}$ & 0.0017 & 0.0031 & 0.638 & 2.26 \\

& &  &  \\ 

\hline
& &  &  \\ 

E15A & 0.0020 &  0.053 & 0.067 &  20.6 & 49.7 \\

& &  &  \\ 

\hline
\hline

\tablenotetext{}{$\beta^{\rm Rot}$ is the ratio of the total rotational energy to the magnitude of the total gravitational energy.
All "$P_0$" models have $R_\Omega=1000$ km (see eq.~\ref{omegaprofile}).\\
Subscripts $i$, $b$, and $500$ imply {\it initial}, {\it at bounce}, and {\it 500 milliseconds after bounce}, respectively.}

\end{tabular}
\end{center}
\end{table}

\figcaption{Initial angular velocity ($\Omega$, rad s$^{-1}$) as a function of radius for the models
considered in this paper.  The solid lines show the $\Omega(r)$ profile imposed on the 
11 M$_\odot$  model of Woosley \& Weaver (1995) with $R_\Omega=1000$ km
and $P_0=1.25$, 2, 3, 4, 5, and 8 seconds (see eq.~\ref{omegaprofile}).
The dashed line shows the angular velocity profile for the model E15A from Heger et al.~(2000).
For simplicity, we refer to the  Woosley \& Weaver (1995)  models by their initial "$P_0$" and 
the Heger et al.~(2000) model as "E15A".
\label{plot:omegap2}}

\figcaption{Neutrino luminosity ($L_\nu$, left panels) in $10^{52}$ erg s$^{-1}$ and
average neutrino energy ($\langle\varepsilon_\nu\rangle$, right panels) in MeV 
as a function of time after bounce for the rotating 11 M$_\odot$ models with $P_0=1.25$, 2, 3, 4, 5, and 8 s (solid lines).  
In all cases, at 200\,ms after bounce, the lowest $L_\nu$ and $\langle\varepsilon_\nu\rangle$ 
correspond to the fastest initial rotation. Note that this hierarchy is preserved
at all times and in all panels except in the case of $L_{\nu_e}$ 400\,ms after bounce
and at breakout.  
Although they cannot be directly compared with the 11 M$_\odot$ models, 
the results for the model E15A from Heger et al.~(2000) are also shown (dashed lines, see Fig.~\ref{plot:omegap2})
The $\nu_e$ breakout burst at $t=0$ (not shown here) is 
largest for the fastest rotator, with $L_{\nu_e}^{\rm peak}\simeq3.1\times10^{53}$ erg s$^{-1}$.  
For both E15A and the model with $P_0=2$ s, $L_{\nu_e}^{\rm peak}\simeq2.6\times10^{53}$ erg s$^{-1}$.  
For the slowest rotator we recover the non-rotating case and 
$L_{\nu_e}^{\rm peak}\simeq2.4\times10^{53}$ erg s$^{-1}$.  See \S\ref{section:rotation} for details.
\label{plot:ltet}}

\figcaption{Mass accretion rate ($\dot{M}$ [M$_\odot$\,s$^{-1}$], upper left panel), 
shear ($d\Omega/d\ln r$) and $\Omega$ ($10^3$ rad s$^{-1}$, upper right panel),
putative saturation magnetic field strength ($\log_{10}[B_{\rm sat}\,\,{\rm G}]$, lower left panel),
and viscous dissipation rate ($\dot{q}_{\rm MRI}$ [erg g$^{-1}$ s$^{-1}$], 
lower right panel) 
versus radius in km for rotating models with 
$P_0=1.25$ (dotted), 2 (dashed), 3 (dot-dashed), and 8\,s (solid)
at $t=105$\,ms after bounce.  These models did not include viscous dissipation.
We show $\dot{q}_{\rm MRI}$ (with $\alpha=0.1$ in $\xi_{\rm MRI}$
as in eq.~\ref{numri}) only for comparison with typical neutrino heating
rates  in models of supernovae ($\dot{q}_\nu$, solid box). 
For clarity of presentation a different radial scale is used
for each panel.
\label{plot:mdshbp}}

\figcaption{The neutrino heating timescale ($\tau_{\rm{H}_\nu}=(P/\rho)/H_\nu$, dot-dashed line),
neutrino cooling timescale ($\tau_{\rm{C}_\nu}=(P/\rho)/C_\nu$, dotted line),
net neutrino heating timescale ($\tau_{\dot{q}_\nu}=(P/\rho)/\dot{q}_\nu$, solid line), and the
advection timescale ($\tau_{\rm Adv}=H/V_r$, dashed line)
in seconds, as a function of radius $r$ (km) at a time 130\,ms
after bounce. 
$R_{\nu_e}$ and $R_{\rm g}$ mark the $\nu_e$ neutrinosphere (for $\varepsilon_{\nu_e}\simeq12$ MeV) and 
the gain radius, respectively.
\label{plot:tcoolp2}}

\figcaption{{\it Left Panel -} Microscopic shear viscosity
of electron neutrinos (solid line), anti-electron neutrinos
(long dashed line), and muon neutrinos (short dashed line)
in the deep interior of a PNS $\sim$105\,ms
after bounce.  For comparison, we show the effective
viscosity due to the MRI (taking $\alpha=0.1$; dotted line), 
which dominates the total neutrino viscosity.
{\it Right Panel -} Damping rate due to neutrino interactions
($\Gamma_{\nu}$) and the optimal growth rate of the MRI ($\Gamma_{{\rm MRI}}\sim\Omega$) in the
semi-transparent regime.  
The Brunt-V\"{a}is\"{a}la frequency $\sqrt{|N^2|}$ is also shown.
\label{plot:nunu2}}

\figcaption{Energy deposition rate as a function of radius in units of 100 MeV baryon$^{-1}$ s$^{-1}$ for
the rotating model with $P_0=2$ seconds, including viscous dissipation via the MRI with $\alpha=0.1$.
Total energy deposition rate ($\dot{q}_{\rm TOT}=\dot{q}_\nu+\dot{q}_{\rm MRI}$, solid line),
viscous energy deposition rate ($\dot{q}_{\rm MRI}$, eq.~\ref{qmri}, dot-dashed line), and neutrino deposition rate 
$\dot{q}_\nu=H_\nu-C_\nu$, long dashed line) are shown.  For comparison, individual contributions
to $\dot{q}_\nu$ from $\nu_e$ (short dashed line) and $\bar{\nu}_e$ (dotted line) are also included.
\label{plot:sinkmp}}

\figcaption{The neutrino heating timescale ($\tau_{\rm{H}_\nu}=(P/\rho)/H_\nu$, dot-short dashed line),
neutrino cooling timescale ($\tau_{\rm{C}_\nu}=(P/\rho)/C_\nu$, dotted line),
net neutrino heating timescale ($\tau_{\dot{q_\nu}}=(P/\rho)/\dot{q}_\nu$, dot-long dashed line), the
advection timescale ($\tau_{\rm Adv}=H/V_r$, long dashed line), the viscous
heating timescale for the MRI ($\tau_{\rm MRI}=(P/\rho)/\dot{q}_{\rm MRI}$, short dashed line),
and the total heating timescale including viscous dissipation 
($\tau_{\dot{q}_{\rm TOT}}=(P/\rho)/(\dot{q}_\nu+\dot{q}_{\rm MRI})$, solid line)
in seconds, as a function of radius $r$ (km) at a time 340\,ms after bounce in our 11 M$_\odot$ progenitor with $P_0=2$ s. 
$R_{\nu_e}$ and $R_{\rm g}$ mark the $\nu_e$ neutrinosphere and gain radius, respectively. The shock radius is $\sim275$ km.
Compare with Figs.~\ref{plot:tcoolp2} and \ref{plot:sinkmp}. Note that exterior to $R_{\rm g}$, as a result of $\dot{q}_{\rm MRI}$,
$\tau_{\rm Adv}$ is greater than $\tau_{\dot{q}_{\rm TOT}}$ (see \S\ref{section:noexp} and \S\ref{section:dissipation}).
\label{plot:tcoolp}}


\figcaption{Angular momentum ($j$, $10^{17}$ cm$^{2}$ s$^{-1}$) as a function of enclosed mass in the
model E15A including viscous dissipation with $\alpha=0.1$ at bounce (solid line, $t=0$), 22 ms (long dashed line), 
100 (dot-short dashed line), 200 (short dashed line), and 350 ms (dot-long dashed line) after bounce.  For reference
the Keplerian angular momentum profile $j_{\rm Kep}=r^2\Omega_{\rm Kep}$ at bounce ($t=0$, dotted line) is also shown. 
\label{plot:angp}}

\figcaption{Radial velocity as a function of radius for various
snapshots in time during the development of explosion in the 11 M$_\odot$ progenitor 
with $P_0=2$ s, including viscous dissipation after bounce with $\alpha=0.1$.
\label{plot:vmrp2}}

\figcaption{Radial position of mass elements as a function of time in a model with $P_0=4$ s
and $\alpha=0.1$.  Viscous dissipation was assumed to occur throughout the radial profile.
The heavy solid lines show the trajectories of the 1.0, 1.2, 1.3, and 1.4 M$_\odot$ mass elements.
At the end of the simulation, the mass cut is at $\simeq1.35$ M$_\odot$.  Note that for clarity
only every tenth mass zone is shown in this figure.
\label{plot:masstee}}

\figcaption{Matter velocity and adiabatic sound speed at two snapshots during explosion in
the model with $P_0=2$ s (compare with Fig.~\ref{plot:vmrp2}).
Note the development of the sonic point on a timescale of $\sim100$ ms, indicating the presence of a transonic
wind.  This neutrino-driven super-sonic outflow is generic to all of our exploding models.
\label{plot:csrp}}

\newpage
\plotone{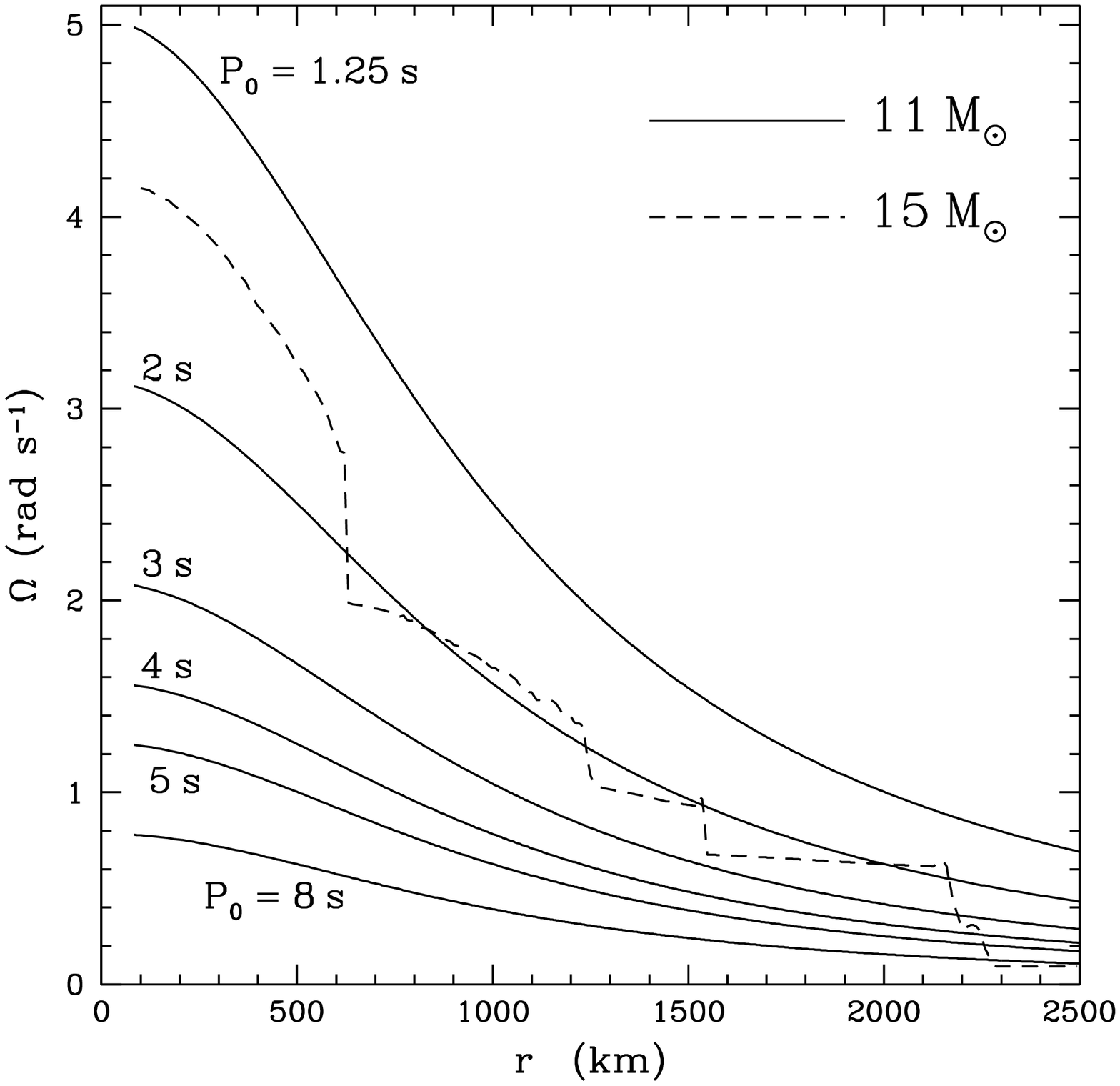}

\newpage
\plotone{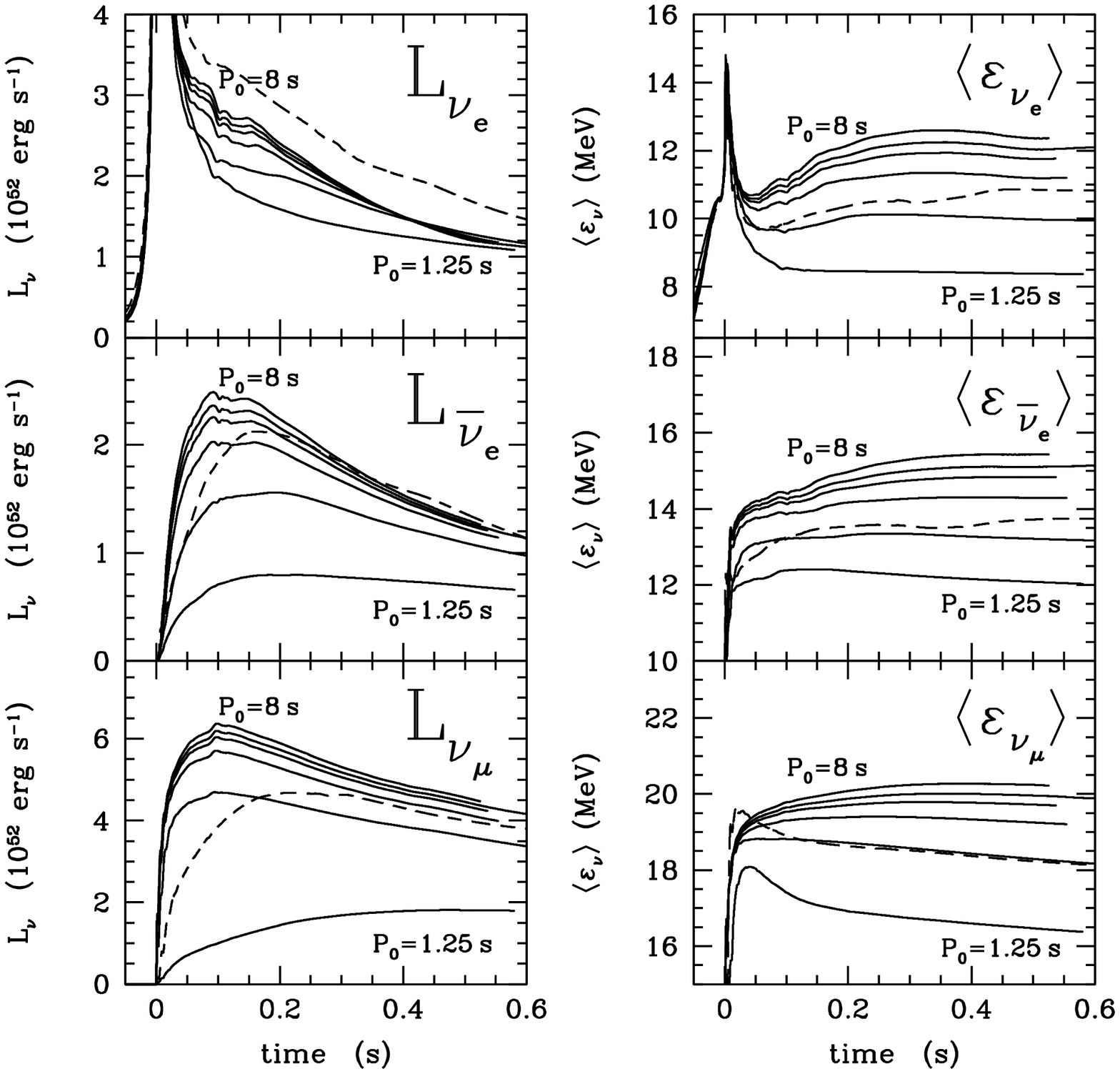}

\newpage
\plotone{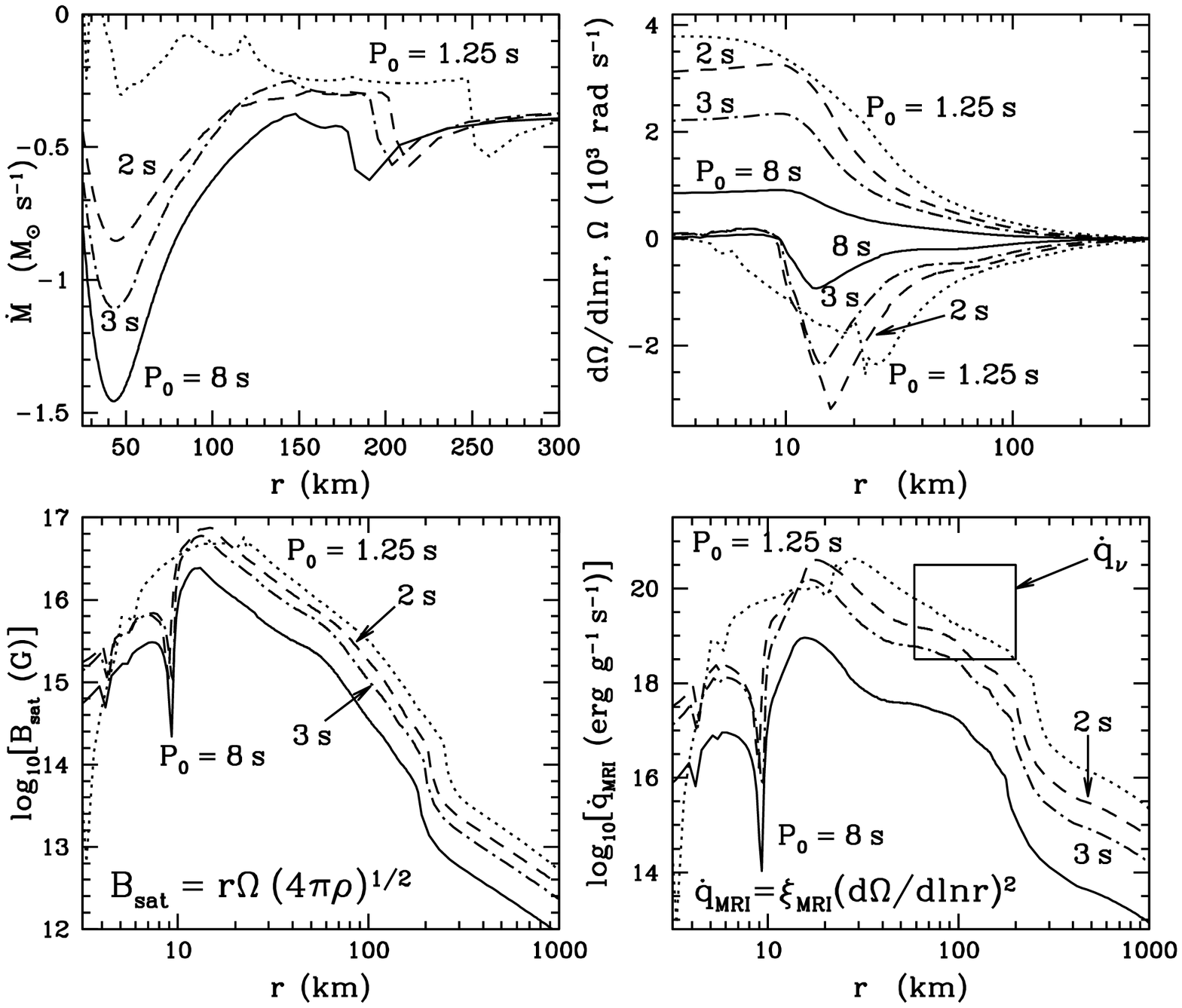}

\newpage
\plotone{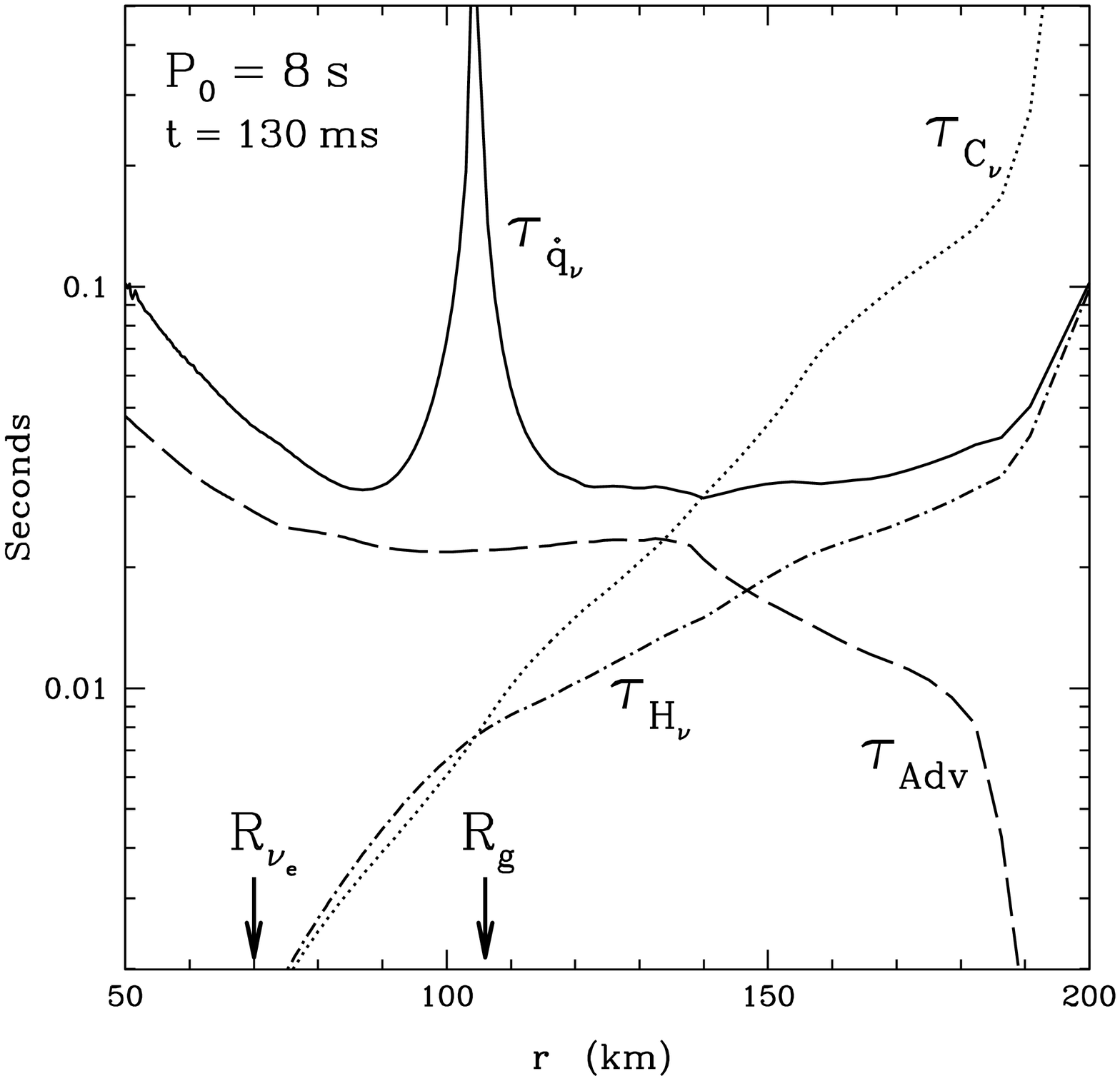}

\newpage
\plotone{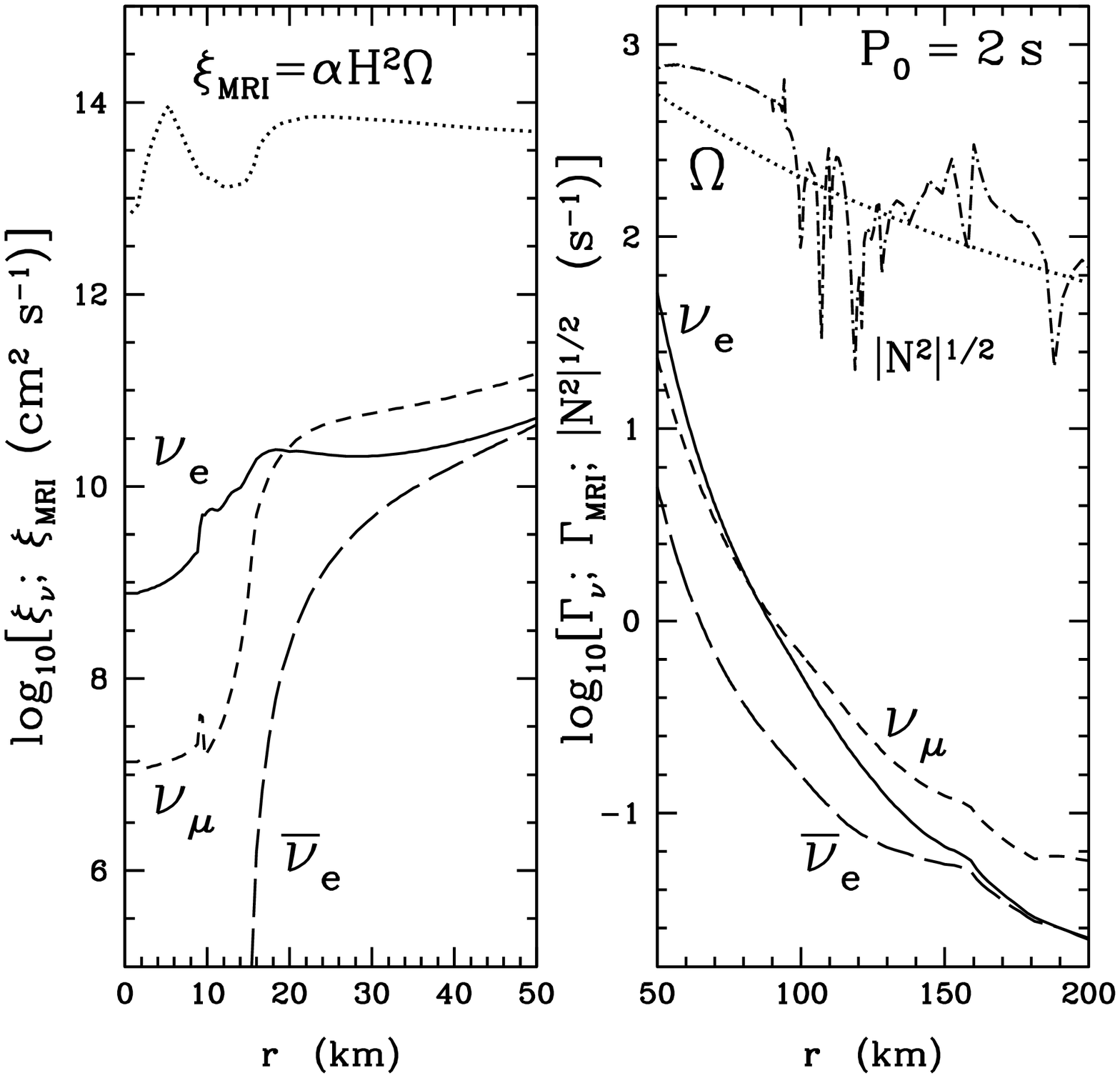}

\newpage
\plotone{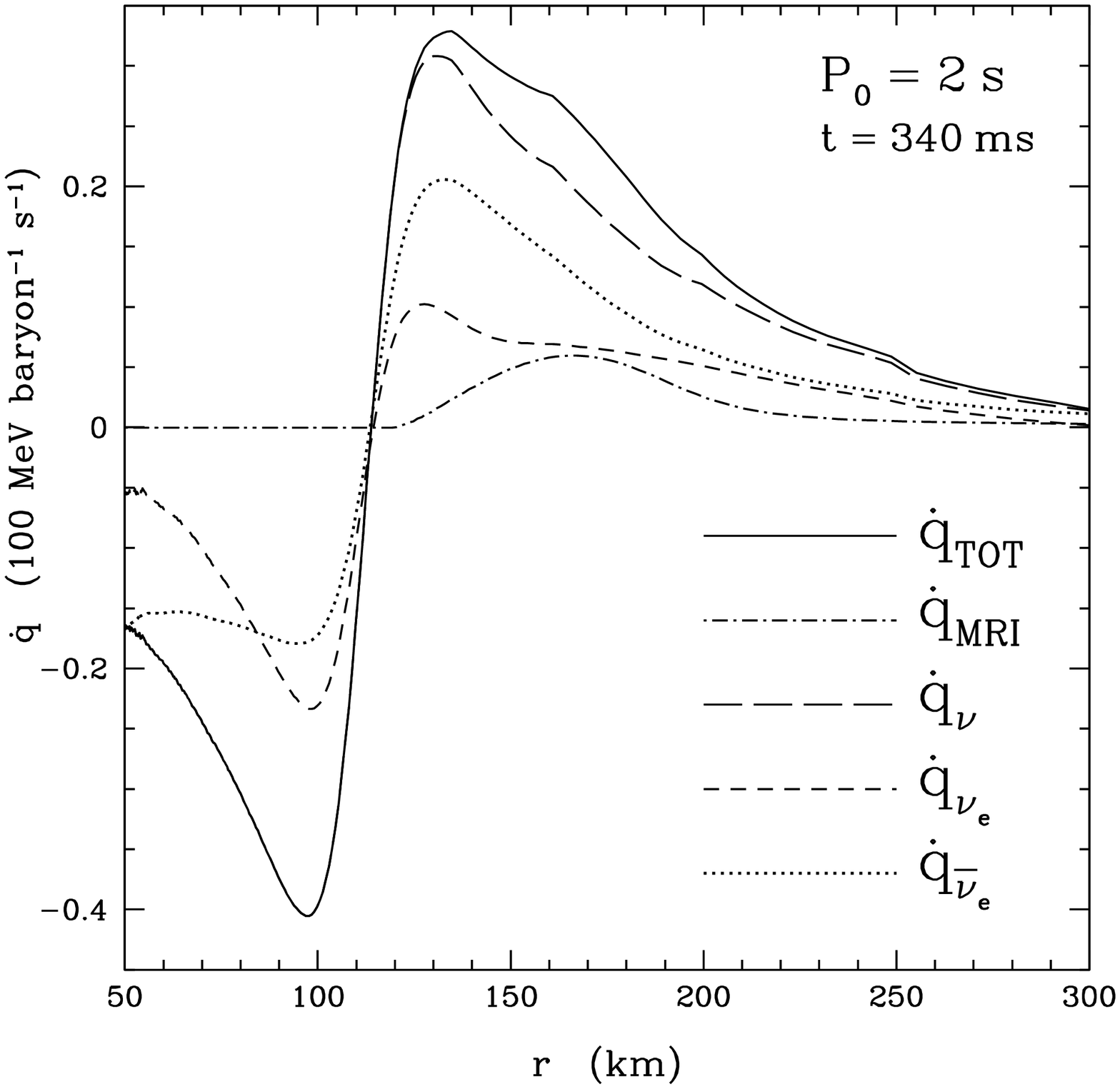}

\newpage
\plotone{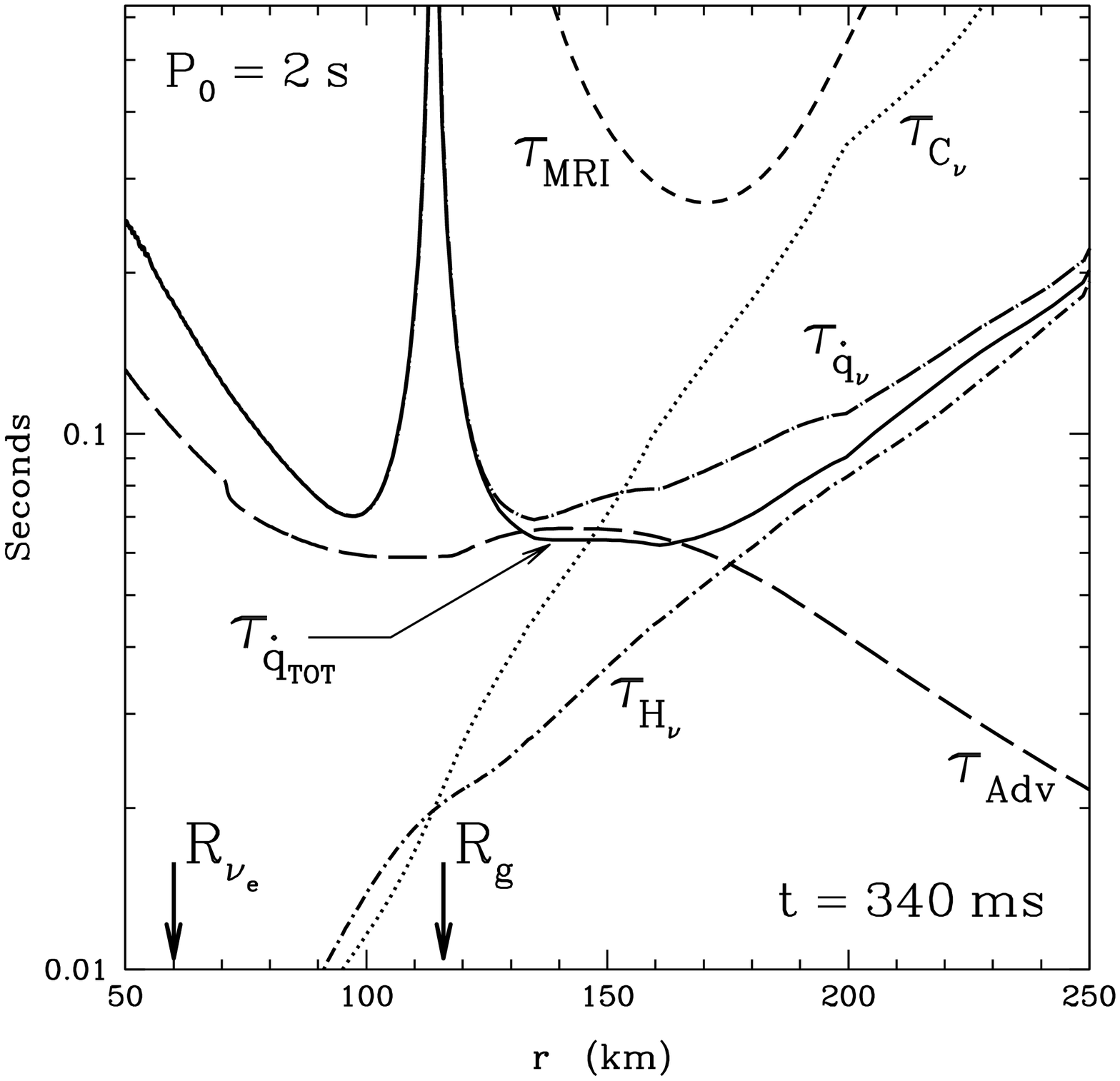}


\newpage
\plotone{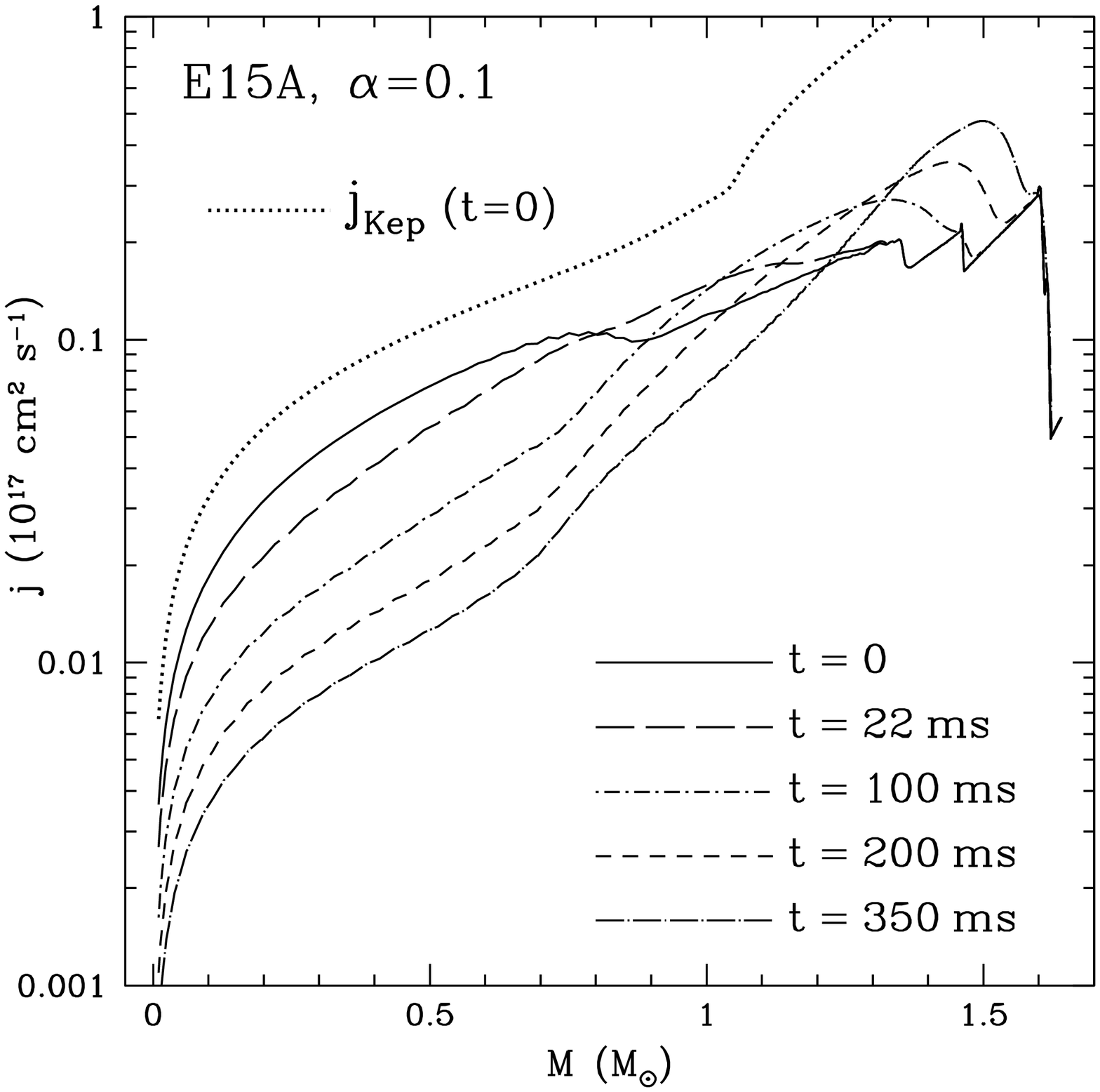}

\newpage
\plotone{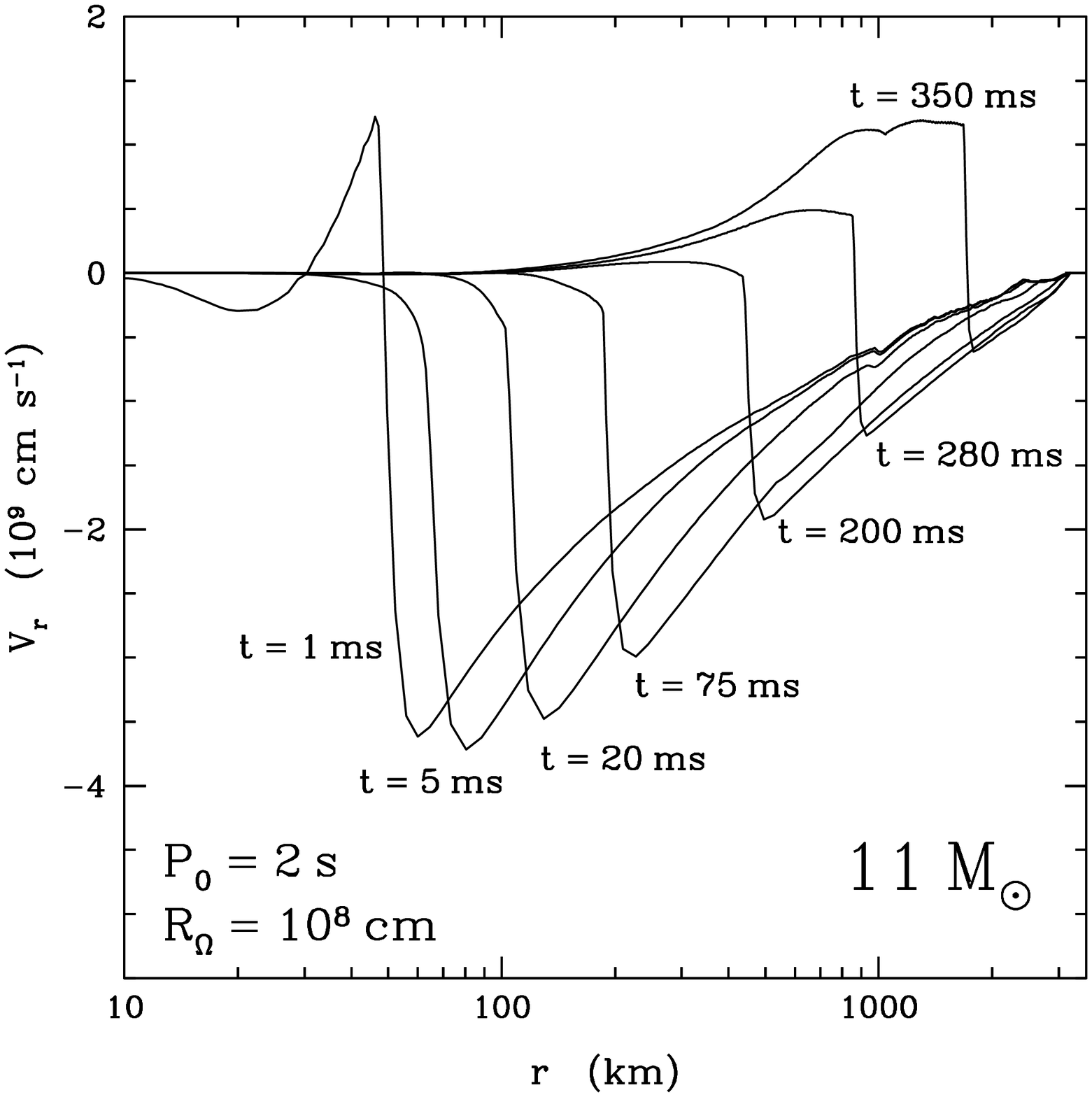}

\newpage
\plotone{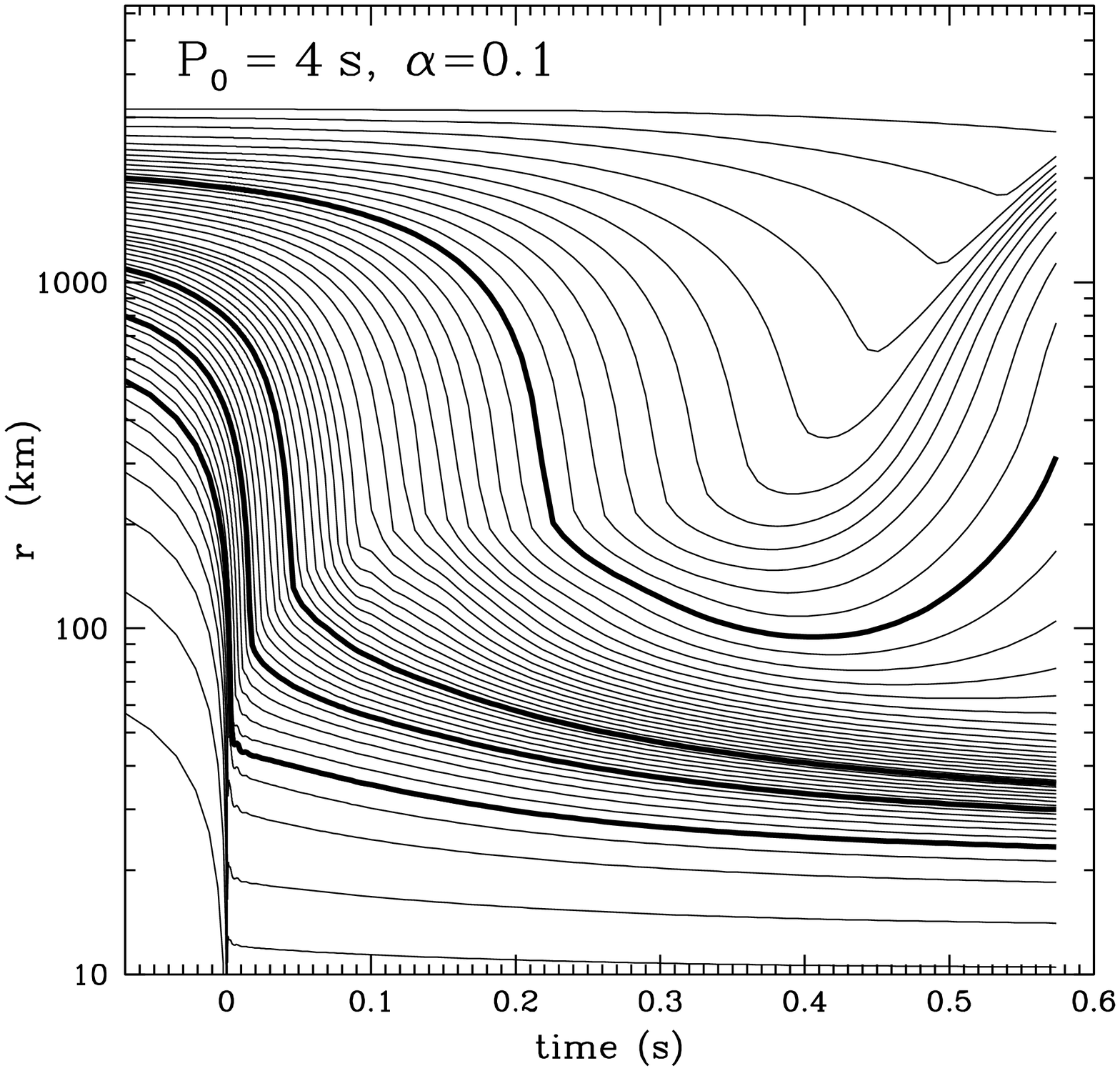}

\newpage
\plotone{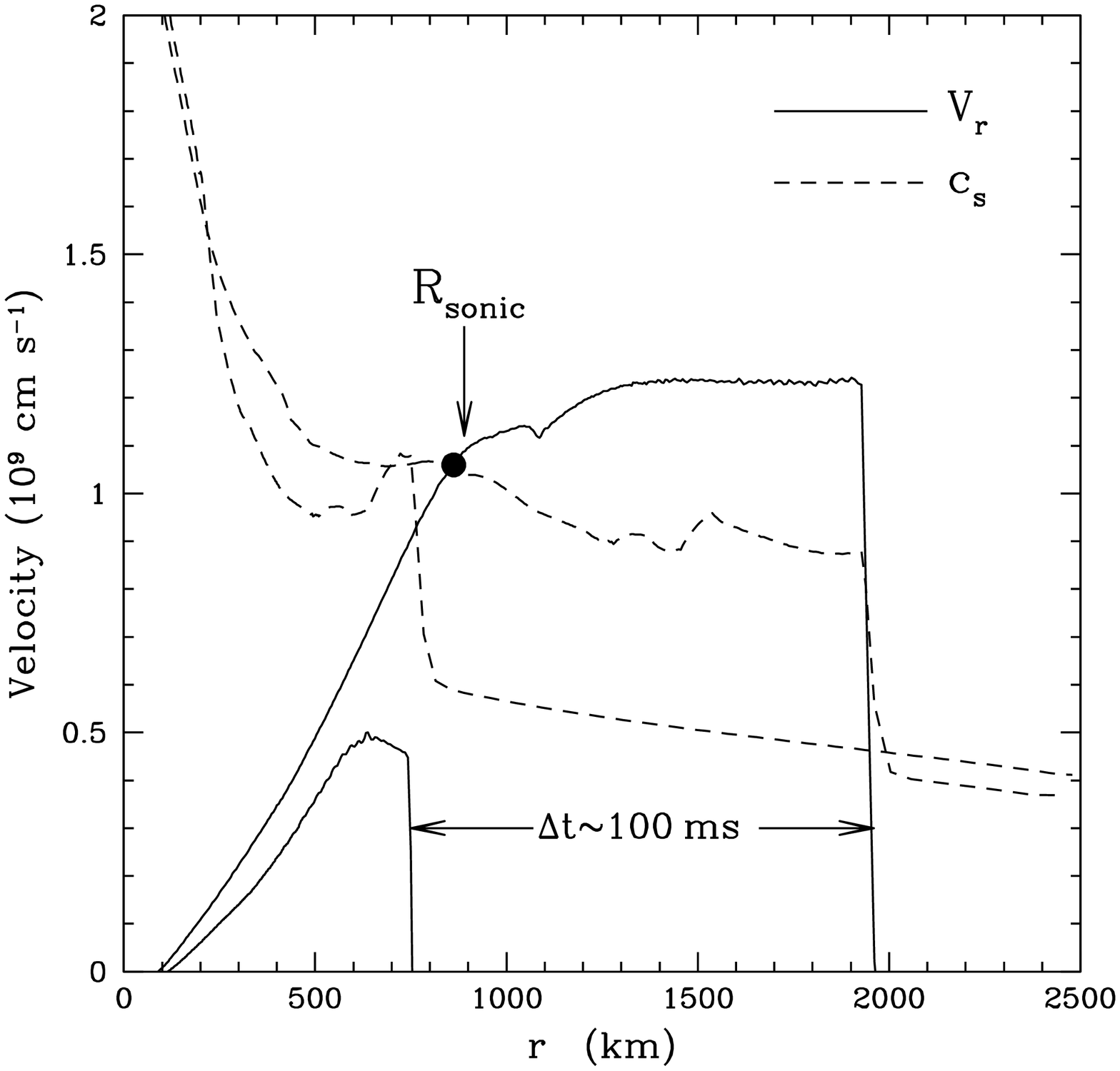}

\end{document}